\documentclass[aps,preprint,nofootinbib,eqsecnum]{revtex4}%
\usepackage{amsmath}
\usepackage{graphicx}
\usepackage{amsfonts}
\usepackage{amssymb}%
\providecommand{\U}[1]{\protect\rule{.1in}{.1in}}
\providecommand{\U}[1]{\protect\rule{.1in}{.1in}}

\begin{document}

\begin{center}
{\leftline {USC-07/HEP-B3 \hfill hep-th/0704.0296}}

{\vskip0.3cm}

{\Large Generalized Twistor Transform And Dualities}

{\Large With A New Description of Particles With Spin }

{\Large Beyond Free and Massless}\footnote{This work was partially supported
by the US Department of Energy, grant number DE-FG03-84ER40168.}

{\vskip0.3cm}

\textbf{Itzhak Bars and Bora Orcal}

{\vskip0.2cm}

\textsl{Department of Physics and Astronomy, }

\textsl{University of Southern California, Los Angeles, CA 90089-0484, USA.}

{\vskip0.5cm}

\textbf{Abstract}
\end{center}

A generalized twistor transform for spinning particles in 3+1 dimensions is
constructed that beautifully unifies many types of spinning systems by mapping
them to the \textit{same twistor} $Z_{A}=\left(
\genfrac{}{}{0pt}{}{\mu^{\dot{\alpha}}}{\lambda_{\alpha}}%
\right)  ,$ thus predicting an infinite set of duality relations
among spinning systems with different Hamiltonians. Usual 1T-physics
is not equipped to explain the duality relationships and unification
between these systems. We use 2T-physics in 4+2 dimensions to
uncover new properties of twistors, and expect that our approach
will prove to be useful for practical applications as well as for a
deeper understanding of fundamental physics. Unexpected structures
for a new description of spinning particles emerge. A unifying
symmetry SU$\left(  2,3\right)  $ that includes conformal symmetry
SU$\left(  2,2\right)  =$SO$\left( 4,2\right)  $ in the massless
case, turns out to be a fundamental property underlying the
dualities of a large set of spinning systems, including those that
occur in high spin theories. This may lead to new forms of string
theory backgrounds as well as to new methods for studying various
corners of M theory. In this paper we present the main concepts, and
in a companion paper we give other details \cite{twistorBO}.

{\vskip0.8cm}

\section{Spinning Particles in 3+1 - Beyond Free and Massless \label{s1}}

The Penrose twistor transform \cite{penrose}-\cite{shirafuji} brings to the
foreground the conformal symmetry SO$\left(  4,2\right)  $ in the dynamics of
massless relativistic particles of any spin in $3+1$ dimensions. The transform
relates the phase space and spin degrees of freedom $x^{\mu},p_{\mu},s^{\mu
\nu}$ to a twistor $Z_{A}=\left(
\genfrac{}{}{0pt}{}{\mu^{\dot{\alpha}}}{\lambda_{\alpha}}%
\right)  $ and reformulates the dynamics in terms of twistors instead of phase
space. The twistor $Z_{A}$ is made up of a pair of SL$\left(  2,C\right)  $
spinors $\mu^{\dot{\alpha}},\lambda_{\alpha},$ $\alpha,\dot{\alpha}=1,2,$ and
is regarded as the 4 components $A=1,2,3,4$ of the Weyl spinor of SO$\left(
4,2\right)  =$SU$\left(  2,2\right)  $.

The well known twistor transform for a spinning massless particle is
\cite{shirafuji}%
\begin{equation}
\mu^{\dot{\alpha}}=-i\left(  \bar{x}+i\bar{y}\right)  ^{\dot{\alpha}\beta
}\lambda_{\beta},\;\;\lambda_{\alpha}\bar{\lambda}_{\dot{\beta}}=p_{\alpha
\dot{\beta}}, \label{penrose0}%
\end{equation}
where $\left(  \bar{x}+i\bar{y}\right)  ^{\dot{\alpha}\beta}=\frac{1}{\sqrt
{2}}\left(  x^{\mu}+iy^{\mu}\right)  \left(  \bar{\sigma}_{\mu}\right)
^{\dot{\alpha}\beta}$ , and $p_{\alpha\dot{\beta}}=\frac{1}{\sqrt{2}}p^{\mu
}\left(  \sigma_{\mu}\right)  _{\alpha\dot{\beta}}$, while $\sigma_{\mu
}=\left(  1,\vec{\sigma}\right)  ,~\bar{\sigma}_{\mu}=\left(  -1,\vec{\sigma
}\right)  $ are Pauli matrices. $x^{\mu}+iy^{\mu}$ is a complexification of
spacetime \cite{penrose}. The helicity $h$ of the particle is determined by
$p\cdot y=h.$ The spin tensor is given by $s^{\mu\nu}=\varepsilon^{\mu\nu
\rho\sigma}y_{\rho}p_{\sigma}$, and it leads to $\frac{1}{2}s^{\mu\nu}%
s_{\mu\nu}=h^{2}.$ The Pauli-Lubanski vector is proportional to the momentum
$W_{\mu}=\frac{1}{2}\varepsilon_{\mu\nu\rho\sigma}s^{\nu\rho}p^{\sigma
}=\left(  y\cdot p\right)  p_{\mu}-p^{2}y^{\mu}=hp_{\mu},$ appropriate for a
massless particle of helicity $h$.

The reformulation of the dynamics in terms of twistors is manifestly
SU$\left(  2,2\right)  $ covariant. It was believed that twistors and the
SO$\left(  4,2\right)  $=SU$\left(  2,2\right)  $ symmetry, interpreted as
conformal symmetry, govern the dynamics of massless particles only, since the
momentum $p^{\mu}$ of the form $p_{\alpha\dot{\beta}}=\lambda_{\alpha}%
\bar{\lambda}_{\dot{\beta}}$ automatically satisfies $p^{\mu}p_{\mu}=0.$

However, recent work has shown that the \textit{same twistor} $Z_{A}=\left(
\genfrac{}{}{0pt}{}{\mu^{\dot{\alpha}}}{\lambda_{\alpha}}%
\right)  $ that describes massless \textit{spinless }particles ($h=0$) also
describes an assortment of other spinless particle dynamical systems
\cite{twistorBP1}\cite{twistorLect}. These include massive and interacting
particles. The mechanism that avoids $p^{\mu}p_{\mu}=0$ \cite{twistorBP1}%
\cite{twistorLect} is explained following Eq.(\ref{z0}) below. The list of
systems includes the following examples worked out explicitly in previous
publications and in unpublished notes.%

\begin{tabular}
[c]{l}%
The massless relativistic particle in $d=4$ flat Minkowski space.\\
The massive relativistic particle in $d=4$ flat Minkowski space.\\
The nonrelativistic free massive particle in 3 space dimensions.\\
The nonrelativistic hydrogen atom (i.e. $1/r$ potential) in 3 space
dimensions.\\
The harmonic oscillator in 2 space dimensions, with its mass $\Leftrightarrow$
an extra dimension.\\
The particle on AdS$_{4}$, or on dS$_{4}.$\\
The particle on AdS$_{3}\times$S$^{1}$ or on $R\times S^{3}.$\\
The particle on AdS$_{2}\times$S$^{2}.$\\
The particle on the Robertson-Walker spacetime.\\
The particle on any maximally symmetric space of positive or negative
curvature.\\
The particle on any of the above spaces modified by any conformal factor.\\
A related family of other particle systems, including some black hole
backgrounds.
\end{tabular}
\newline In this paper we will discuss these for the case of $d=4$
\textit{with spin }($h\neq0$). It must be emphasized that while the phase
spaces (and therefore dynamics, Hamiltonian, etc.) in these systems are
different, the twistors $\left(  \mu^{\dot{\alpha}},\lambda_{\alpha}\right)  $
are the same. For example, the massive particle phase space $\left(  x^{\mu
},p^{\mu}\right)  _{massive}$ and the one for the massless particle $\left(
x^{\mu},p^{\mu}\right)  _{massless}$ are not the same $\left(  x^{\mu},p^{\mu
}\right)  ,$ rather they can be obtained from one another by a non-linear
transformation for any value of the mass parameter $m$ \cite{twistorBP1}, and
similarly, for all the other spaces mentioned above. However, under such
\textquotedblleft duality\textquotedblright\ transformations from one system
to another, the twistors for all the cases are the same up to an overall phase
transformation
\begin{equation}
\left(  \mu^{\dot{\alpha}},\lambda_{\alpha}\right)  _{massive}=\left(
\mu^{\dot{\alpha}},\lambda_{\alpha}\right)  _{massless}=\cdots=\left(
\mu^{\dot{\alpha}},\lambda_{\alpha}\right)  .
\end{equation}

This unification also shows that all of these systems share the same
SO$\left(  4,2\right)  $=SU$\left(  2,2\right)  $ global symmetry of the
twistors. This SU$\left(  2,2\right)  $ is interpreted as conformal symmetry
for the massless particle phase space, but has other meanings as a hidden
symmetry of all the other systems in their own phase spaces. Furthermore, in
the quantum physical Hilbert space, the symmetry is realized in the
\textit{same unitary representation} of SU$\left(  2,2\right)  ,$ with the
same Casimir eigenvalues (see (\ref{twistingCasimirsa},\ref{twistingCasimirb})
below), for all the systems listed above.

The underlying reason for such fantastic looking properties cannot be found in
one-time physics (1T-physics) in 3+1 dimensions, but is explained in two-time
physics (2T-physics) \cite{2treviews} as being due to a local Sp$\left(
2,R\right)  $ symmetry. The Sp$\left(  2,R\right)  $ symmetry which acts in
phase space makes position and momentum indistinguishable at any instant and
requires one extra space and one extra time dimensions to implement it, thus
showing that the unification relies on an underlying spacetime in 4+2
dimensions. It was realized sometime ago that in 2T-physics twistors emerge as
a gauge choice \cite{2ttwistor}, while the other systems are also gauge
choices of the same theory in 4+2 dimensions. The 4+2 phase space can be gauge
fixed to many 3+1 phase spaces that are distinguishable from the point of view
of 1T-physics, without any Kaluza-Klein remnants, and this accounts for the
different Hamiltonians that have a duality relationship with one another. We
will take advantage of the properties of 2T-physics to build the general
twistor transform that relates these systems including spin.

Given that the field theoretic formulation of 2T-physics in 4+2 dimensions
yields the Standard Model of Particles and Forces in 3+1 dimensions as a gauge
choice \cite{2tstandardM}, including spacetime supersymmetry \cite{susy2t},
and given that twistors have simplified QCD computations \cite{witten2}%
\cite{cachazo}, we expect that our twistor methods will find useful applications.

\section{Twistor Lagrangian}

The Penrose twistor description of massless spinning particles requires that
the pairs $\left(  \mu^{\dot{\alpha}},i\bar{\lambda}_{\dot{\alpha}}\right)  $
or their complex conjugates $\left(  \lambda_{\alpha},i\bar{\mu}^{\alpha
}\right)  $ be canonical conjugates and satisfy the helicity constraint given
by%
\begin{equation}
\bar{Z}^{A}Z_{A}=\bar{\lambda}_{\dot{\alpha}}\mu^{\dot{\alpha}}+\bar{\mu
}^{\alpha}\lambda_{\alpha}=2h. \label{helicitycons}%
\end{equation}
Indeed, Eq.(\ref{penrose0}) satisfies this property provided $y\cdot p=h$.
Here we have defined the \textbf{\={4} }of SU$\left(  2,2\right)  $ as the
contravariant twistor
\begin{equation}
\bar{Z}^{A}\equiv\left(  Z^{\dagger}\eta_{2,2}\right)  ^{A}=\left(
\bar{\lambda}_{\dot{\alpha}}\;\bar{\mu}^{\alpha}\right)  ,\;\;\eta
_{2,2}=\left(
\genfrac{}{}{0pt}{}{0}{1}%
\genfrac{}{}{0pt}{}{1}{0}%
\right)  =\text{SU}\left(  2,2\right)  ~\text{metric.} \label{contra}%
\end{equation}
The canonical structure, along with the constraint $\bar{Z}^{A}Z_{A}=2h$
follows from the following worldline action for twistors
\begin{equation}
S_{h}=\int d\tau\left[  i\bar{Z}_{A}\left(  D_{\tau}Z^{A}\right)  -2h\tilde
{A}\right]  ,\;\;D_{\tau}Z^{A}\equiv\frac{\partial Z^{A}}{\partial\tau
}-i\tilde{A}Z^{A}. \label{action}%
\end{equation}
In the case of $h=0$ it was shown that this action emerges as a gauge choice
of a more general action in 2T-physics \cite{twistorBP1}\cite{twistorLect}.
Later in the paper, in Eq.(\ref{actionV}) we give the $h\neq0$ 2T-physics
action from which (\ref{action}) is derived as a gauge choice. The derivative
part of this action gives the canonical structure $S_{0}=\int d\tau i\bar
{Z}_{A}\left(  \partial_{\tau}Z^{A}\right)  =i\int d\tau\left[  \bar{\lambda
}_{\dot{\alpha}}\partial_{\tau}\mu^{\dot{\alpha}}+\bar{\mu}^{\alpha}%
\partial_{\tau}\lambda_{\alpha}\right]  $ that requires $\left(  \mu
^{\dot{\alpha}},i\bar{\lambda}_{\dot{\alpha}}\right)  $ or their complex
conjugates $\left(  \lambda_{\alpha},i\bar{\mu}^{\alpha}\right)  $ to be
canonical conjugates. The 1-form $\tilde{A}d\tau$ is a U$\left(  1\right)  $
gauge field on the worldline, $D_{\tau}Z^{A}$ is the U$\left(  1\right)  $
gauge covariant derivative that satisfies $\delta_{\varepsilon}\left(
D_{\tau}Z^{A}\right)  =i\varepsilon\left(  D_{\tau}Z^{A}\right)  $ for
$\delta_{\varepsilon}\tilde{A}=\partial\varepsilon/\partial\tau$ and
$\delta_{\varepsilon}Z^{A}=i\varepsilon Z^{A}$. The term $2h\tilde{A}$ is
gauge invariant since it transforms as a total derivative under the
infinitesimal gauge transformation$.$ $2h\tilde{A}$ was introduced in
\cite{twistorBP1}\cite{twistorLect} as being an integral part of the twistor
formulation of the spinning particle action.

Our aim is to show that this action describes not only massless spinning
particles, but also all of the other particle systems listed above
\textit{with spin}. This will be done by constructing the twistor transform
from $Z_{A}$ to the phase space and spin degrees of freedom of these systems,
and claiming the unification of dynamics via the generalized twistor
transform. This generalizes the work of \cite{twistorBP1}\cite{twistorLect}
which was done for the $h=0$ case of the action in (\ref{action}). We will use
2T-physics as a tool to construct the general twistor transform, so this
unification is equivalent to the unification achieved in 2T-physics.

\section{Massless Particle With Any Spin in 3+1 Dimensions}

In our quest for the general twistor transform with spin, we first
discuss an alternative to the well known twistor transform of
Eq.(\ref{penrose0}). Instead of the $y^{\mu}\left(  \tau\right)  $
that appears in the complexified spacetime $x^{\mu}+iy^{\mu}$ we
introduce an SL$\left(  2,C\right)  $ \textit{bosonic}\footnote{This
is similar to the \textit{fermionic} case in \cite{shirafuji}. The
bosonic spinor $v$ can describe any spin $h$.} spinor
$v^{\dot{\alpha}}\left(  \tau\right)  $ and its complex conjugate
$\bar {v}^{\alpha}\left(  \tau\right)  ,$ and write the general
vector $y^{\mu}$ in the matrix form as
$y^{\dot{\alpha}\beta}=hv^{\dot{\alpha}}\bar{v}^{\beta }+\omega
p^{\dot{\alpha}\beta},$ where $\omega\left(  \tau\right)  $ is an
arbitrary gauge freedom that drops out. Then the helicity condition
$y\cdot p=h$ takes the form $\bar{v}pv=1.$ Furthermore, we can write
$\lambda_{\alpha }=p_{\alpha\dot{\beta}}v^{\dot{\beta}}$ since this
automatically satisfies
$\lambda_{\alpha}\bar{\lambda}_{\dot{\beta}}=p_{\alpha\dot{\beta}}$
when $p^{2}=\left(  \bar{v}pv-1\right)  =0$ are true. With this
choice of
variables, the Penrose transform of Eq.(\ref{penrose0}) takes the new form%
\begin{equation}
\lambda_{\alpha}=\left(  pv\right)  _{\alpha},\;\mu^{\dot{\alpha}}=\left[
\left(  -i\bar{x}p+h\right)  v\right]  ^{\dot{\alpha}},\;p^{2}=\left(  \bar
{v}pv-1\right)  =0, \label{new}%
\end{equation}
where the last equation is a set of constraints on the degrees of freedom
$x^{\mu},p_{\mu},v^{\dot{\alpha}},\bar{v}^{\alpha}.$

If we insert the twistor transform (\ref{new}) into the action (\ref{action}),
the twistor action turns into the action for the phase space and spin degrees
of freedom $x^{\mu},p_{\mu},v^{\dot{\alpha}},\bar{v}^{\alpha}$%
\begin{equation}
S_{h}=\int d\tau\left\{  \dot{x}^{\mu}p_{\mu}-\frac{e}{2}p^{2}+ih\left[
\left(  \bar{v}p\right)  D_{\tau}v-\overline{D_{\tau}v}\left(  pv\right)
\right]  -2h\tilde{A}\right\}  . \label{actionps}%
\end{equation}
where $D_{\tau}v=\dot{v}-i\tilde{A}v$ is the U$\left(  1\right)  $ gauge
covariant derivative and we have included the Lagrange multiplier $e$ to
impose $p^{2}=0$ when we don't refer to twistors. The equation of motion for
$\tilde{A}$ imposes the second constraint $\bar{v}pv-1=0$ that implies
U$\left(  1\right)  $ gauge invariance\footnote{If this action is taken
without the U$\left(  1\right)  $ constraint $\left(  \tilde{A}=0\right)  $,
then the excitations in the $v$ sector describe an infinite tower of massless
states with all helicities from zero to infinity (here we rescale $\sqrt
{2h}v\rightarrow v$)
\begin{equation}
S_{all~spins}=\int d\tau\left\{  \dot{x}^{\mu}p_{\mu}-\frac{e}{2}p^{2}%
+\frac{i}{2}\left[  \bar{v}p\dot{v}-\overline{\dot{v}}pv\right]  \right\}
\label{all}%
\end{equation}
The spectrum coincides with the spectrum of the infinite slope limit of string
theory with all helicities $\frac{1}{2}\bar{v}pv$. This action has a hidden
SU$\left(  2,3\right)  $ symmetry that includes SU$\left(  2,2\right)  $
conformal symmetry. This is explained in the rest of the paper by the fact
that this action is a gauge fixed version of a 2T-physics master action
(\ref{actionV},\ref{actionV2}) in 4+2 dimensions with manifest SU$\left(
2,3\right)  $ symmetry. A related approach has been pursued also in
\cite{fedo1}-\cite{fedo4} in 3+1 dimensions in the context of only massless
particles. Along with the manifestly SU$(2,3)$ symmetric 2T-physics actions,
we are proposing here a unified 2T-physics setting for discussing high spin
theories \cite{vasiliev} including all the dual versions of the high spin
theories related to the spinning physical systems listed in section
(\ref{s1}). \label{alll}}.

From the global Lorentz symmetry of (\ref{actionps}), the Lorentz generator is
computed via Noether's theorem $J^{\mu\nu}=x^{\mu}p^{\nu}-x^{\nu}p^{\mu
}+s^{\mu\nu},$ with $s^{\mu\nu}=\frac{i}{2}h\bar{v}\left(  p\bar{\sigma}%
^{\mu\nu}+\sigma^{\mu\nu}p\right)  v.$ The helicity is determined by computing
the Pauli-Lubanski vector $W^{\mu}=\frac{1}{2}\varepsilon^{\mu\nu\lambda
\sigma}s_{\nu\lambda}p_{\sigma}=\left(  h\bar{v}pv\right)  p^{\mu}.$ The
helicity operator $h\bar{v}pv$ reduces to the constant $h$ in the U$\left(
1\right)  $ gauge invariant sector.

The action (\ref{actionps}) gives a description of a massless particle with
any helicity $h$ in terms of the SL$\left(  2,C\right)  $ bosonic spinors
$v,\bar{v}.$ We note its similarity to the standard superparticle action
\cite{ibhanson}\cite{brink} written in the first order formalism. The
difference with the superparticle is that the \textit{fermionic} spacetime
spinor $\theta^{\dot{\alpha}}$ of the superparticle is replaced with the
\textit{bosonic} spacetime spinor $v^{\dot{\alpha}},$ and the gauge field
$\tilde{A}$ imposes the U$\left(  1\right)  $ gauge symmetry constraint
$\bar{v}pv-1=0$ that restricts the system to a single, but arbitrary helicity
state given by $h$.

Just like the superparticle case, our action has a local \textit{kappa
symmetry} with a \textit{bosonic} local spinor parameter $\kappa_{\alpha
}\left(  \tau\right)  $, namely%
\begin{align}
\delta_{\kappa}v^{\dot{\alpha}}  &  =\bar{p}^{\dot{\alpha}\beta}\kappa_{\beta
},\;\delta_{\kappa}x_{\mu}=\frac{ih}{\sqrt{2}}\left(  \left(  \delta_{\kappa
}\bar{v}\right)  \sigma_{\mu}v-\bar{v}\sigma_{\mu}\left(  \delta_{\kappa
}v\right)  \right)  ,\;\label{kappa1}\\
\delta_{\kappa}p^{\mu}  &  =0,\;\;\delta_{\kappa}e=-ih~\left[  \bar{\kappa
}\left(  D_{\tau}v\right)  -\left(  \overline{D_{\tau}v}\right)
\kappa\right]  ,\;\delta_{\kappa}\tilde{A}=0. \label{kappa2}%
\end{align}
These kappa transformations mix the phase space degrees of freedom $\left(
x,p\right)  $ with the spin degrees of freedom $v,\bar{v}.$ The
transformations $\delta_{\kappa}x_{\mu},\delta_{\kappa}e$ are non-linear$.$

Let us count \textit{physical} degrees of freedom. By using the kappa and the
$\tau$-reparametrization symmetries one can choose the lightcone gauge. From
phase space $x^{\mu},p^{\mu}$ there remains 3 positions and 3 momentum degrees
of freedom. One of the two complex components of $v^{\dot{\alpha}}$ is set to
zero by using the kappa symmetry, so $v^{\dot{\alpha}}=\left(
\genfrac{}{}{0pt}{}{v}{0}%
\right)  $. The phase of the remaining component is eliminated by choosing the
U$\left(  1\right)  $ gauge, and finally its magnitude is fixed by solving the
constraint $\bar{v}pv-1=0$ to obtain $v^{\dot{\alpha}}=\left(  p^{+}\right)
^{-1/2}\left(
\genfrac{}{}{0pt}{}{1}{0}%
\right)  .$ Therefore, there are no \textit{independent} physical degrees of
freedom in $v$. The remaining degrees of freedom for the particle of any spin
are just the three positions and momenta, and the constant $h$ that appears in
$s^{\mu\nu}.$ This is as it should be, as seen also by counting the physical
degrees of freedom from the twistor point of view. When we consider the other
systems listed in the first section, we should expect that they too are
described by the same number of degrees of freedom since they will be obtained
from the same twistor, although they obey different dynamics (different
Hamiltonians) in their respective phase spaces.

The lightcone quantization of the the massless particle systems described by
the actions (\ref{actionps},\ref{all}) is performed after identifying the
physical degrees of freedom as discussed above. The lightcone quantum spectrum
and wavefunction are the expected ones for spinning \textit{massless}
particles, and agree with their covariant quantization given in
\cite{fedo1}-\cite{fedo5}.

\section{2T-physics With Sp$\left(  2,R\right)  ,$ SU$(2,3)$ and Kappa
Symmetries}

The similarity of (\ref{actionps}) to the action of the superparticle provides
the hint for how to lift it to the 2T-physics formalism, as was done for the
superparticle \cite{super2t}\cite{2ttwistor} and the twistor superstring
\cite{2tsuperstring}\cite{2tstringtwistors}. This requires lifting 3+1 phase
space $\left(  x^{\mu},p_{\mu}\right)  $ to 4+2 phase space $\left(
X^{M},P_{M}\right)  $ and lifting the SL$\left(  2,C\right)  $ spinors
$v,\bar{v}$ to the SU$\left(  2,2\right)  $ spinors $V_{A},\bar{V}^{A}.$ The
larger set of degrees of freedom $X^{M},P_{M},V_{A},\bar{V}^{A}$ that are
covariant under the global symmetry SU$\left(  2,2\right)  =$SO$\left(
4,2\right)  ,$ include gauge degrees of freedom, and are subject to gauge
symmetries and constraints that follow from them as described below.

The point is that the SU$\left(  2,2\right)  $ invariant constraints on
$X^{M},P_{M},V_{A},\bar{V}^{A}$ have a wider set of solutions than just the
3+1 system of Eq.(\ref{actionps}) we started from. This is because 3+1
dimensional spin \& phase space has many different embeddings in 4+2
dimensions, and those are distinguishable from the point of view of 1T-physics
because target space \textquotedblleft time\textquotedblright\ and
corresponding \textquotedblleft Hamiltonian\textquotedblright\ are different
in different embeddings, thus producing the different dynamical systems listed
in section (\ref{s1}). The various 1T-physics solutions are reached by simply
making gauge choices. One of the gauge choices for the action we give below in
Eq.(\ref{actionV}) is the twistor action of Eq.(\ref{action}). Another gauge
choice is the 4+2 spin \& phase space action in terms of the lifted spin \&
phase space $X^{M},P_{M},V_{A},\bar{V}^{A}$ as given in Eq.(\ref{actionV2}).
The latter can be further gauge fixed to produce all of the systems listed in
section (\ref{s1}) including the action (\ref{actionps}) for the massless
spinning particle with any spin. All solutions still remember that there is a
hidden \textit{global} symmetry SU$\left(  2,2\right)  =$SO$\left(
4,2\right)  ,$ so all systems listed in section (\ref{s1}) are realizations of
the same unitary representation of SU$\left(  2,2\right)  $ whose Casimir
eigenvalues will be given below.

For the $4+2$ version of the superparticle \cite{super2t} that is similar to
the action in (\ref{actionV2}), this program was taken to a higher level in
\cite{2ttwistor} by embedding the fermionic supercoordinates in the coset of
the supergroup SU$\left(  2,2|1\right)  /$SU$\left(  2,2\right)  \times
$U$\left(  1\right)  $. We will follow the same route here, and embed the
bosonic SU$\left(  2,2\right)  $ spinors $V_{A},\bar{V}^{A}$ in the left coset
SU$\left(  2,3\right)  /$SU$\left(  2,2\right)  \times$U$\left(  1\right)  .$
This coset will be regarded as the gauging of the group SU$\left(  2,3\right)
$ under the subgroup $\left[  \text{SU}\left(  2,2\right)  \times
\text{U}\left(  1\right)  \right]  _{L}$ from the left side. Thus the most
powerful version of the action that reveals the global and gauge symmetries is
obtained when it is organized in terms of the $X_{i}^{M}\left(  \tau\right)
$, $g\left(  \tau\right)  $ and $\tilde{A}\left(  \tau\right)  $ degrees of
freedom described as%
\[%
\begin{array}
[c]{l}%
\text{\textit{4+2 phase space} }\left(
\genfrac{}{}{0pt}{}{X^{M}\left(  \tau\right)  }{P_{M}\left(  \tau\right)  }%
\right)  \equiv X_{i}^{M}\left(  \tau\right)  ,\text{ }i=1,2,~\text{\textit{
doublets of} Sp}\left(  2,R\right)  \text{ \textit{gauge symmetry,}}\\
\text{\textit{group element} }g\left(  \tau\right)  \subset\text{SU}\left(
2,3\right)  \text{ \textit{subject to} [SU}\left(  2,2\right)  \times
\text{U}\left(  1\right)  \text{]}_{L}\times\text{U}\left(  1\right)
_{L+R}\text{ \textit{gauge symmetry.}}%
\end{array}
\]
We should mention that the $h=0$ version of this theory, and the corresponding
twistor property, was discussed in \cite{twistorBP1}, by taking $g\left(
\tau\right)  \subset$SU$\left(  2,2\right)  $ and dropping all of the
U$\left(  1\right)  $'s. So, the generalized theory that includes spin has the
new features that involves SU$\left(  2,2\right)  \rightarrow$SU$\left(
2,3\right)  $ and the U$\left(  1\right)  $ structures. The action has the
following form%
\begin{equation}
S_{h}=\int d\tau\left\{  \frac{1}{2}\varepsilon^{ij}\left(  D_{\tau}X_{i}%
^{M}\right)  X_{j}^{N}\eta_{MN}+Tr\left(  \left(  iD_{\tau}g\right)
g^{-1}\left(
\genfrac{}{}{0pt}{}{\mathcal{L}}{0}%
\genfrac{}{}{0pt}{}{0}{0}%
\right)  \right)  -2h\tilde{A}\right\}  , \label{actionV}%
\end{equation}
where $\varepsilon^{ij}=\left(
\genfrac{}{}{0pt}{}{0}{-1}%
\genfrac{}{}{0pt}{}{1}{0}%
\right)  ^{ij}$ is the antisymmetric Sp$\left(  2,R\right)  $ metric, and
$D_{\tau}X_{i}^{M}=\partial_{\tau}X_{i}^{M}-A_{i}^{~j}X_{j}^{M}$ is the
Sp$\left(  2,R\right)  $ gauge covariant derivative, with the 3 gauge
potentials $A^{ij}=\varepsilon^{ik}A_{k}^{~j}=\left(
\genfrac{}{}{0pt}{}{A}{C}%
\genfrac{}{}{0pt}{}{C}{B}%
\right)  .$ For SU$\left(  2,3\right)  $ the group element is pseudo-unitary,
$g^{-1}=\left(  \eta_{2,3}\right)  g^{\dagger}\left(  \eta_{2,3}\right)
^{-1},$ where $\eta_{2,3}$ is the SU$\left(  2,3\right)  $ metric $\eta
_{2,3}=\left(
\genfrac{}{}{0pt}{}{\eta_{2,2}}{0}%
\genfrac{}{}{0pt}{}{0}{-1}%
\right)  .$ The covariant derivative $D_{\tau}g$ is given by%
\begin{equation}
D_{\tau}g=\partial_{\tau}g-i\tilde{A}\left[  q,g\right]  ,\;\;q=\frac{1}%
{5}\left(
\begin{tabular}
[c]{l|l}%
$1_{4\times4}$ & $~0$\\\hline
$~~0$ & $-4$%
\end{tabular}
\ \ \right)  \label{Dg}%
\end{equation}
where the generator of U$\left(  1\right)  _{L+R}$ is proportional to the
5$\times$5 traceless matrix $q\in~$u$\left(  1\right)  \in~$su$\left(
2,3\right)  _{L+R}.$ The last term of the action $-2h\tilde{A},$ which is also
the last term of the action (\ref{action}), is invariant under the U$\left(
1\right)  _{L+R}$ since it transforms to a total derivative. Finally, the
$4\times4$ traceless matrix $\left(  \mathcal{L}\right)  _{A}^{~B}\in
$su$\left(  2,2\right)  \in$su$\left(  2,3\right)  $ that appears on the
\textit{left side} of $g$ (or right side of $g^{-1}$) is
\begin{equation}
\left(  \mathcal{L}\right)  _{A}^{~B}\equiv\left(  \frac{1}{4i}\Gamma
_{MN}\right)  _{A}^{~B}~L^{MN},\;\;L^{MN}=\varepsilon^{ij}X_{i}^{M}X_{j}%
^{N}=X^{M}P^{N}-X^{N}P^{M}.
\end{equation}
where $\Gamma_{MN}=\frac{1}{2}\left(  \Gamma_{M}\bar{\Gamma}_{N}-\Gamma
_{N}\bar{\Gamma}_{M}\right)  $ are the 4$\times4$ gamma-matrix representation
of the 15 generators of SU$\left(  2,2\right)  $. A detailed description of
these gamma matrices is given in \cite{susy2t}.

The symmetries of actions of this type for any group or supergroup $g$ were
discussed in \cite{2ttwistor}\cite{2tsuperstring}\cite{2tstringtwistors}%
\cite{twistorLect}. The only modification of that discussion here is due to
the inclusion of the U$\left(  1\right)  $ gauge field $\tilde{A}.$ In the
absence of the $\tilde{A}$ coupling the global symmetry is given by the
transformation of $g\left(  \tau\right)  $ from the right side $g\left(
\tau\right)  \rightarrow g\left(  \tau\right)  g_{R}$ where $g_{R}\subset
$SU$(2,3)_{R}.$ However, in our case, the presence of the coupling with the
U$\left(  1\right)  _{L+R}$ charge $q$ breaks the global symmetry down to the
(SU$(2,2)\times$U$\left(  1\right)  $)$_{R}$ subgroup that acts on the right
side of $g$.

So the global symmetry is given by
\begin{equation}
\text{global:\ \ }g\left(  \tau\right)  \rightarrow g\left(  \tau\right)
h_{R},\;h_{R}\in\lbrack\text{SU}(2,2)\times\text{U}\left(  1\right)
]_{R}\subset\text{SU}(2,3)_{R}.
\end{equation}
Using Noether's theorem we deduce the conserved global charges as the
$[$SU$(2,2)\times$U$\left(  1\right)  ]_{R}$ components of the the following
SU$(2,3)_{R}$ Lie algebra valued matrix $J_{\left(  2,3\right)  }$
\begin{equation}
J_{\left(  2,3\right)  }=g^{-1}\left(
\genfrac{}{}{0pt}{}{\mathcal{L}}{0}%
\genfrac{}{}{0pt}{}{0}{0}%
\right)  g=\left(
\genfrac{}{}{0pt}{}{\mathcal{J+}\frac{1}{4}J_{0}}{-\bar{j}}%
\genfrac{}{}{0pt}{}{j}{-J_{0}}%
\right)  ,\;\;J_{2,3}=\eta_{2,3}\left(  J_{2,3}\right)  ^{\dagger}\left(
\eta_{2,3}\right)  ^{-1}, \label{conserved}%
\end{equation}
The traceless $4\times4$ matrix $\left(  \mathcal{J}\right)  _{A}^{~B}%
=\frac{1}{4i}\Gamma^{MN}J_{MN}$ is the conserved SU$(2,2)=$SO$\left(
4,2\right)  $ charge and $J_{0}$ is the conserved U$\left(  1\right)  $
charge. Namely, by using the equations of motion one can verify $\partial
_{\tau}\left(  \mathcal{J}\right)  _{A}^{~B}=0$ and $\partial_{\tau}J_{0}=0.$
The spinor charges $j_{A},\bar{j}^{A}$ are not conserved\footnote{In the high
spin version of (\ref{actionV}) with $\tilde{A}=0$, the global symmetry is
SU$\left(  2,3\right)  _{R}$ and $j_{A},\bar{j}^{A}$ are conserved.} due to
the coupling of $\tilde{A}.$ As we will find out later in Eq.(\ref{general2}),
$j_{A}$ is proportional to the twistor%
\begin{equation}
j_{A}=\sqrt{J_{0}}Z_{A}\text{,} \label{rho}%
\end{equation}
up to an irrelevant gauge transformation. It is important to note that
$\mathcal{J}$ and $J_{0}$ are invariant on shell under the gauge symmetries
discussed below. Therefore they generate physical symmetries $[$%
SU$(2,2)\times$U$\left(  1\right)  ]_{R}$ under which all gauge invariant
physical states are classified.

The local symmetries of this action are summarized as
\begin{equation}
\text{Sp}\left(  2,R\right)  \times\left(
\begin{array}
[c]{cc}%
\text{SU}\left(  2,2\right)  & \frac{3}{4}kappa\\
\frac{3}{4}kappa & \text{U}\left(  1\right)
\end{array}
\right)  _{left} \label{locals}%
\end{equation}
The Sp$\left(  2,R\right)  $ is manifest in (\ref{actionV}). The rest
corresponds to making local SU$\left(  2,3\right)  $ transformations on
$g\left(  \tau\right)  $ from the left side $g\left(  \tau\right)  \rightarrow
g_{L}\left(  \tau\right)  g\left(  \tau\right)  ,$ as well as transforming
$X_{i}^{M}=\left(  X^{M},P^{M}\right)  $ as vectors with the local subgroup
SU$\left(  2,2\right)  _{L}=$SO$\left(  4,2\right)  ,$ and $A^{ij}$ under the
kappa. The $3/4$ kappa symmetry which is harder to see will be discussed in
more detail below. These symmetries coincide with those given in previous
discussions in \cite{2ttwistor}\cite{2tsuperstring}\cite{2tstringtwistors}%
\cite{twistorLect} despite the presence of $\tilde{A}$. The reason is that the
U$\left(  1\right)  _{L+R}$ covariant derivative $D_{\tau}g$ in Eq.(\ref{Dg})
can be replaced by a purely U$\left(  1\right)  _{R}$ covariant derivative
$D_{\tau}g=\partial_{\tau}g+igq\tilde{A}$ because the difference drops out in
the trace in the action (\ref{actionV}). Hence the symmetries on left side of
$g\left(  \tau\right)  \rightarrow g_{L}\left(  \tau\right)  g\left(
\tau\right)  $ remain the same despite the coupling of $\tilde{A}.$

We outline the roles of each of these local symmetries. The Sp$\left(
2,R\right)  $ gauge symmetry can reduce $X^{M},P_{M}$ to any of the phase
spaces in 3+1 dimensions listed in section (\ref{s1}). This is the same as the
$h=0$ case discussed in \cite{twistorBP1}. The [SU$\left(  2,2\right)  \times
$U$\left(  1\right)  $]$_{L}$ gauge symmetry can reduce $g\left(  \tau\right)
\subset$SU$\left(  2,3\right)  $ to the coset $g\rightarrow t\left(  V\right)
\in$SU$\left(  2,3\right)  /[$SU$\left(  2,2\right)  \times$U$\left(
1\right)  ]_{L}$ parameterized by the SU$\left(  2,2\right)  \times$U$\left(
1\right)  $ spinors $\left(  V_{A},\bar{V}^{A}\right)  $ as shown in
Eq.(\ref{tV}). The remaining $3/4$ kappa symmetry, whose action is shown in
Eq.(\ref{kappa}), can remove up to 3 out of the 4 parameters in the $V_{A}$.
The U$\left(  1\right)  _{L+R}$ symmetry can eliminate the phase of the
remaining component in $V$. Finally the constraint due to the equation of
motion of $\tilde{A}$ fixes the magnitude of $V$. In terms of counting, there
remains only 3 position and 3 momentum \textit{physical} degrees of freedom,
plus the constant $h,$ in agreement with the counting of physical degrees of
freedom of the twistors.

It is possible to gauge fix the symmetries (\ref{locals}) partially to exhibit
some intermediate covariant forms. For example, to reach the SL$\left(
2,C\right)  $ covariant massless particle described by the action
(\ref{actionps}) from the 2T-physics action above, we take the massless
particle gauge by using two out of the three Sp$\left(  2,R\right)  $ gauge
parameters to rotate the $M=+^{\prime}$ doublet to the form $\left(
\genfrac{}{}{0pt}{}{X^{+^{\prime}}}{P^{+^{\prime}}}%
\right)  \left(  \tau\right)  =\left(
\genfrac{}{}{0pt}{}{1}{0}%
\right)  ,$ and solving explicitly two of the Sp$\left(  2,R\right)  $
constraints $X^{2}=X\cdot P=0$
\begin{equation}
X^{M}=(\overset{+^{\prime}}{1},\;\overset{-^{\prime}}{\;\frac{x^{2}%
\mathbf{\;}}{2\mathbf{\;}}}~,~~\overset{\mu}{x^{\mu}}\left(  \tau\right)
),\;\;\;P^{M}=(\overset{+^{\prime}}{0},\;\overset{-^{\prime}}{\;x\cdot
p}~,~~\overset{\mu}{p^{\mu}}\left(  \tau\right)  ). \label{massless}%
\end{equation}
This is the same as the $h=0$ massless case in \cite{twistorBP1}. There is a
tau reparametrization gauge symmetry as a remnant of Sp$\left(  2,R\right)  .$
Next, the [SU$\left(  2,2\right)  \times$U$\left(  1\right)  $]$_{L}$ gauge
symmetry reduces $g\left(  \tau\right)  \rightarrow t\left(  V\right)  $
written in terms of $\left(  V_{A},\bar{V}^{A}\right)  $ as given in
Eq.(\ref{tV})$,$ and the $3/4$ kappa symmetry reduces the SU$\left(
2,2\right)  $ spinor $V_{A}\rightarrow\left(
\genfrac{}{}{0pt}{}{v^{\dot{\alpha}}}{0}%
\right)  $ to the two components SL$\left(  2,C\right)  $ doublet
$v^{\dot{\alpha}},$ with a leftover kappa symmetry as discussed in
Eqs.(\ref{kappa1}-\ref{kappa2}). The gauge fixed form of $g$ is then
\begin{equation}
g=\exp\left(
\begin{tabular}
[c]{cc|c}%
$0$ & $0$ & $\sqrt{2h}v^{\dot{\alpha}}$\\
$0$ & $0$ & $0$\\\hline
$0$ & $\sqrt{2h}\bar{v}^{\alpha}$ & $0$%
\end{tabular}
\right)  =\left(
\begin{tabular}
[c]{cc|c}%
$1$ & $hv^{\dot{\alpha}}\bar{v}^{\beta}$ & $\sqrt{2h}v^{\dot{\alpha}}$\\
$0$ & $1$ & $0$\\\hline
$0$ & $\sqrt{2h}\bar{v}^{\alpha}$ & $1$%
\end{tabular}
\right)  \in\text{SU}\left(  2,3\right)  . \label{massless2}%
\end{equation}
The inverse $g^{-1}=\left(  \eta_{2,3}\right)  g^{\dagger}\left(  \eta
_{2,3}\right)  ^{-1}$ is given by replacing $v,\bar{v}$ by $\left(  -v\right)
,\left(  -\bar{v}\right)  .$ Inserting the gauge fixed forms of $X,P,g$
(\ref{massless},\ref{massless2}) into the action (\ref{actionV}) reduces it to
the massless spinning particle action (\ref{actionps}). Furthermore, inserting
these $X,P,g$ into the expression for the current in (\ref{conserved}) gives
the conserved SU$\left(  2,2\right)  $ charges $\mathcal{J}$ (see
Eqs.(\ref{J},\ref{Vfixed})) which have the significance of the hidden
conformal symmetry of the gauge fixed action (\ref{actionps}). This hidden
symmetry is far from obvious in the form (\ref{actionps}), but it is
straightforward to derive from the 2T-physics action as we have just outlined.

Partial or full gauge fixings of (\ref{actionV}) similar to (\ref{massless}%
,\ref{massless2}) produce the actions, the hidden SU$\left(  2,2\right)  $
symmetry, and the twistor transforms with spin of all the systems listed in
section (\ref{s1}). These were discussed for $h=0$ in \cite{twistorBP1}, and
we have now shown how they generalize to any spin $h\neq0,$ with further
details below. It is revealing, for example, to realize that the
\textit{massive} spinning particle has a hidden SU$\left(  2,2\right)  $
\textquotedblleft mass-deformed conformal symmetry\textquotedblright,
including spin, not known before, and that its action can be reached by gauge
fixing the action (\ref{actionV}), or by a twistor transform from
(\ref{action}). The same remarks applied to all the other systems listed in
section (\ref{s1}) are equally revealing. For more information see our related
paper \cite{twistorBO}.

Through the gauge (\ref{massless},\ref{massless2}), the twistor transform
(\ref{new}), and the massless particle action (\ref{actionps}), we have
constructed a bridge between the manifestly SU$\left(  2,2\right)  $ invariant
twistor action (\ref{action}) for any spin and the 2T-physics action
(\ref{actionV}) for any spin. This bridge will be made much more transparent
in the following sections by building the general twistor transform.

\section{2T-physics Action with $X^{M},P^{M},V_{A},\bar{V}^{A}$ in 4+2
Dimensions}

We have hinted above that there is an intimate relation between the 2T-physics
action (\ref{actionV}) and the twistor action (\ref{action}). In fact the
twistor action is just a gauged fixed version of the more general 2T-physics
action (\ref{actionV}). Using the \textit{local} SU$\left(  2,2\right)
=$SO$\left(  4,2\right)  $ and local Sp$\left(  2,R\right)  $ symmetries of
the general action (\ref{actionV}) we can rotate $X^{M}\left(  \tau\right)
,P^{M}\left(  \tau\right)  $ to the following form that also solves the
Sp$\left(  2,R\right)  $ constraints $X_{i}\cdot X_{j}=$ $X^{2}=P^{2}=X\cdot
P=0$ \cite{twistorBP1}\cite{twistorLect}%
\begin{equation}
X^{M}=(\overset{+^{\prime}}{1},\;\overset{-^{\prime}}{\;0}~,~~\overset{+}%
{0},\;\overset{-}{\;0}~,~~\overset{i}{0}),\;\;\;P^{M}=(\overset{+^{\prime}}%
{0},\;\overset{-^{\prime}}{\;0}~,~~\overset{+}{1},\;\overset{-}{\;0}%
~,~~\overset{i}{0}). \label{XPconstant}%
\end{equation}
This completely eliminates all phase space degrees of freedom. We are left
with the gauge fixed action $S_{h}=\int d\tau\left\{  Tr\left(  \frac{1}%
{2}\left(  D_{\tau}g\right)  g^{-1}\left(
\genfrac{}{}{0pt}{}{\Gamma^{-^{\prime}-}}{0}%
\genfrac{}{}{0pt}{}{0}{0}%
\right)  \right)  -2h\tilde{A}\right\}  ,$ where $\left(  i\mathcal{L}\right)
\rightarrow\frac{1}{2}\Gamma^{-^{\prime}-}L^{+^{\prime}-^{\prime}},$ and
$L^{+^{\prime}-^{\prime}}=1$. Due to the many zero entries in the 4$\times4$
matrix $\Gamma^{-^{\prime}-}$ \cite{twistorBP1}, only one column from $g$ in
the form $\left(
\genfrac{}{}{0pt}{}{Z_{A}}{Z_{5}}%
\right)  $and one row from $g^{-1}$ in the form $\left(  \bar{Z}^{A},-\bar
{Z}_{5}\right)  $ can contribute in the trace, and therefore the action
becomes $S_{h}=\int d\tau\left\{  i\bar{Z}^{A}\dot{Z}_{A}-i\bar{Z}_{5}\dot
{Z}_{5}+\tilde{A}\left(  \bar{Z}_{5}Z_{5}-2h\right)  \right\}  .$ Here
$\bar{Z}_{5}\dot{Z}_{5}$ drops out as a total derivative since the magnitude
of the complex number $Z_{5}$ is a constant $\bar{Z}_{5}Z_{5}=2h$.
Furthermore, we must take into account $\bar{Z}^{A}Z_{A}-\bar{Z}_{5}Z_{5}=0$
which is an off-diagonal entry in the matrix equation $g^{-1}g=1.$ Then we see
that the 2T-physics action (\ref{actionV}) reduces to the twistor action
(\ref{action}) with the gauge choice (\ref{XPconstant})\footnote{In the high
spin version of (\ref{actionV}) without $\tilde{A}$ (see footnote
(\ref{alll})), we replace $Z_{5}=e^{i\phi}\sqrt{\bar{Z}^{A}Z_{A}}$ and after
dropping a total derivative, the twistor equivalent becomes $S_{all~spins}%
=\int d\tau\left\{  i\bar{Z}^{A}\dot{Z}_{A}+\bar{Z}Z\dot{\phi}\right\}  $. For
a more covariant version that displays the SU$\left(  2,3\right)  $ global
symmetry, we introduce a new U$\left(  1\right)  $ gauge field for the overall
phase of $\left(
\genfrac{}{}{0pt}{}{Z_{A}}{Z_{5}}%
\right)  $ and write $S_{all~spins}=\int d\tau\left\{  i\bar{Z}^{A}\dot{Z}%
_{A}-i\bar{Z}_{5}\dot{Z}_{5}+\tilde{B}\left(  \bar{Z}^{A}Z_{A}-\bar{Z}%
_{5}Z_{5}\right)  \right\}  .$\label{alltwist}}.

Next let us gauge fix the 2T-physics action (\ref{actionV}) to a manifestly
SU$\left(  2,2\right)  =$SO$\left(  4,2\right)  $ invariant version in flat
4+2 dimensions, in terms of the phase space \& spin degrees of freedom
$X^{M},P^{M},V_{A},\bar{V}^{A}$. For this we use the $[$SU$(2,2)\times
$U$(1)]_{left}$ symmetry to gauge fix $g$
\begin{equation}
\text{gauge fix:\ \ }g\rightarrow t\left(  V\right)  \in\frac{\text{SU}%
(2,3)}{[\text{SU}(2,2)\times\text{U}(1)]_{left}}%
\end{equation}
The coset element $t\left(  V\right)  $ is parameterized by the SU$(2,2)$
spinor $V$ and its conjugate $\bar{V}=V^{\dagger}\eta_{2,2}$ and given by the
5$\times$5 SU$(2,3)$ matrix\footnote{Arbitrary fractional powers of the matrix
$\left(  1-2hV\bar{V}\right)  $ are easily computed by expanding in a series
and then resuming to obtain $\left(  1-2hV\bar{V}\right)  ^{\gamma}=1+$
$V\bar{V}\left(  \left(  1-2h\bar{V}V\right)  ^{\gamma}-1\right)  /\bar{V}V.$
\label{powers}}%
\begin{equation}
t\left(  V\right)  =\left(
\begin{array}
[c]{cc}%
\left(  1-2hV\bar{V}\right)  ^{-1/2} & 0\\
0 & \left(  1-2h\bar{V}V\right)  ^{-1/2}%
\end{array}
\right)  \left(
\begin{array}
[c]{cc}%
1 & \sqrt{2h}V\\
\sqrt{2h}\bar{V} & 1
\end{array}
\right)  . \label{tV}%
\end{equation}
The factor $2h$ is inserted for a convenient normalization of $V.$ Note that
the first matrix commutes with the second one, so it can be written in either
order. The inverse of the group element is $t^{-1}\left(  V\right)  =\left(
\eta_{2,3}\right)  t^{\dagger}\left(  \eta_{2,3}\right)  ^{-1}=t\left(
-V\right)  ,$ as can be checked explicitly $t\left(  V\right)  t\left(
-V\right)  =1.$ Inserting this gauge in (\ref{actionV}) the action becomes%
\begin{align}
S_{h}  &  =\int d\tau\left\{  \dot{X}\cdot P-\frac{1}{2}A^{ij}X_{i}\cdot
X_{j}-\frac{1}{2}\Omega^{MN}L_{MN}-2h\tilde{A}\left(  \frac{\bar{V}%
\mathcal{L}V}{1-2h\bar{V}V}-1\right)  \right\} \label{actionV2}\\
&  =\int d\tau\left\{  \frac{1}{2}\varepsilon^{ij}\left(  \hat{D}_{\tau}%
X_{i}^{M}\right)  X_{j}^{N}\eta_{MN}-2h\tilde{A}\left(  \frac{\bar
{V}\mathcal{L}V}{1-2h\bar{V}V}-1\right)  \right\}  \label{actionV3}%
\end{align}
where
\begin{equation}
\hat{D}_{\tau}X_{i}^{M}=\partial_{\tau}X_{i}^{M}-A_{i}^{~j}X_{j}^{M}%
-\Omega^{MN}X_{iN} \label{covder}%
\end{equation}
is a covariant derivative for local Sp$\left(  2,R\right)  $ as well as local
SU$\left(  2,2\right)  =$SO$\left(  4,2\right)  $ but with a composite
SO$\left(  4,2\right)  $ connection $\Omega^{MN}\left(  V\left(  \tau\right)
\right)  $ given conveniently in the following forms
\begin{equation}
\frac{1}{2}\Omega^{MN}\Gamma_{MN}=\left[  \left(  i\partial_{\tau}t\right)
t^{-1}\right]  _{SU\left(  2,2\right)  },\;\text{or~~}\frac{1}{2}\Omega
^{MN}L_{MN}=-Tr\left(  \left(  i\partial_{\tau}t\right)  t^{-1}\left(
\genfrac{}{}{0pt}{}{\mathcal{L}}{0}%
\genfrac{}{}{0pt}{}{0}{0}%
\right)  \right)  . \label{omega1}%
\end{equation}
Thus, $\Omega$ is the SU$\left(  2,2\right)  $ projection of the SU$\left(
2,3\right)  $ Cartan connection and given explicitly as
\begin{equation}
\frac{1}{2}\Omega^{MN}\Gamma_{MN}=2h\frac{\left(  \dot{V}-V\frac{\bar{V}%
\dot{V}}{\bar{V}V}\right)  \bar{V}-V\left(  \overline{\dot{V}}-\bar{V}%
\frac{\overline{\dot{V}}V}{\bar{V}V}\right)  }{\sqrt{1-2h\bar{V}V}\left(
1+\sqrt{1-2h\bar{V}V}\right)  }+h\left(  \frac{V\bar{V}}{\bar{V}V}-\frac{1}%
{4}\right)  \frac{\bar{V}\dot{V}-\overline{\dot{V}}V}{\left(  1-2h\bar
{V}V\right)  } \label{omega}%
\end{equation}

The action (\ref{actionV2},\ref{actionV3}) is manifestly invariant under
global SU$\left(  2,2\right)  =$SO$\left(  4,2\right)  $ rotations, and under
local $U\left(  1\right)  $ phase transformations applied on $V_{A},\bar
{V}^{A}$. The conserved global symmetry currents $\mathcal{J}$ and $J_{0}$ can
be derived either directly from (\ref{actionV2}) by using Noether's theorem,
or by inserting the gauge fixed form of $g\rightarrow t\left(  V\right)  $
into Eq.(\ref{conserved})\footnote{In the high spin version ($\tilde{A}=0$)
the conserved charges include $j_{A}$ as part of SU$\left(  2,3\right)  _{R}$
global symmetry. It is then also convenient to rescale $\sqrt{2h}V\rightarrow
V$ in Eqs.(\ref{tV}-\ref{J0}) to eliminate an irrelevant constant.}
$J_{\left(  2,3\right)  }=t^{-1}\left(
\genfrac{}{}{0pt}{}{\mathcal{L}}{0}%
\genfrac{}{}{0pt}{}{0}{0}%
\right)  t$%
\begin{align}
\mathcal{J}  &  \mathcal{=}\frac{1}{\sqrt{1-2hV\bar{V}}}\mathcal{L}\frac
{1}{\sqrt{1-2hV\bar{V}}}-\frac{1}{4}J_{0},\;\;J_{0}=\frac{2h\bar{V}%
\mathcal{L}V}{1-2h\bar{V}V}\label{J}\\
j_{A}  &  =\sqrt{2h}\frac{1}{\sqrt{1-2hV\bar{V}}}\mathcal{L}V\frac{1}%
{\sqrt{1-2h\bar{V}V}} \label{J0}%
\end{align}
According to the equation of motion for $\tilde{A}$ that follows from the
action (\ref{actionV2}) we must have the following constraint (this means
U$\left(  1\right)  $ gauge invariant physical sector)%
\begin{equation}
\frac{\bar{V}\mathcal{L}V}{1-2h\bar{V}V}=1. \label{constraintVL}%
\end{equation}
Therefore, in the physical sector the conserved $[$SU$(2,2)\times
$U$(1)]_{right}$ charges take the form
\begin{equation}
\text{physical sector: }J_{0}=2h,\;\;\mathcal{J=}\frac{1}{\sqrt{1-2hV\bar{V}}%
}\mathcal{L}\frac{1}{\sqrt{1-2hV\bar{V}}}-\frac{h}{2}. \label{physical}%
\end{equation}

Let us now explain the local kappa symmetry of the action (\ref{actionV2}%
,\ref{actionV3}). The action (\ref{actionV2}) is still invariant under the
bosonic local $3/4$ kappa symmetry inherited from the action (\ref{actionV}).
The kappa transformations of $g\left(  \tau\right)  $ in the general action
(\ref{actionV2}) correspond to local coset elements $\exp\left(
\genfrac{}{}{0pt}{}{0}{\bar{K}}%
\genfrac{}{}{0pt}{}{K}{0}%
\right)  \in$SU$(2,3)_{left}/[$SU$(2,2)\times$U$(1)]_{left}$ with a special
form of the spinor $K_{A}$
\begin{equation}
K_{A}=X_{i}\cdot\left(  \Gamma\kappa^{i}\left(  \tau\right)  \right)
_{A}=X_{M}\left(  \Gamma^{M}\kappa^{1}\right)  _{A}+P_{M}\left(  \Gamma
^{M}\kappa^{2}\right)  _{A}, \label{kappaform}%
\end{equation}
with $\kappa^{iA}\left(  \tau\right)  $ two arbitrary local
spinors\footnote{In this special form only 3 out of the 4 components of
$K_{A}$ are effectively independent gauge parameters. This can be seen easily
in the special frame for $X^{M},P^{M}$ given in Eq.(\ref{XPconstant}).
\label{effective}}. Now that $g$ has been gauge fixed $g\rightarrow t\left(
V\right)  $, the kappa transformation must be taken as the naive kappa
transformation on $g$ followed by a [SU$\left(  2,2\right)  \times$U$\left(
1\right)  ]_{left}$ gauge transformation which restores the gauge fixed form
of $t\left(  V\right)  $
\begin{equation}
t\left(  V\right)  \rightarrow t\left(  V^{\prime}\right)  =\left[
\exp\left(
\genfrac{}{}{0pt}{}{-\omega}{0}%
\genfrac{}{}{0pt}{}{0}{Tr\left(  \omega\right)  }%
\right)  \right]  \left[  \exp\left(
\genfrac{}{}{0pt}{}{0}{\bar{K}}%
\genfrac{}{}{0pt}{}{K}{0}%
\right)  \right]  t\left(  V\right)
\end{equation}
The SU$\left(  2,2\right)  $ part of the restoring gauge transformation must
also be applied on $X^{M},P^{M}$. Performing these steps we find the
infinitesimal version of this transformation \cite{super2t}$\;$
\begin{equation}
\delta_{\kappa}V=\frac{1}{\sqrt{1-2hV\bar{V}}}K\frac{1}{\sqrt{1-2h\bar{V}V}%
},\;\;\delta_{\kappa}X_{i}^{M}=~\omega^{MN}X_{iN},\;\;\delta_{\kappa}%
A^{ij}=\text{see below,} \label{kappa}%
\end{equation}
where $\omega^{MN}\left(  K,V\right)  ~$has the same form as $\Omega^{MN}$ in
Eq.(\ref{omega}) but with $\dot{V}$ replaced by the $\delta_{\kappa}V$ given
above. The covariant derivative $\hat{D}_{\tau}X_{i}^{M}$ in Eq.(\ref{covder})
is covariant under the local SU$\left(  2,2\right)  $ transformation with
parameter $\omega^{MN}\left(  K,V\right)  $ (this is best seen from the
projected Cartan connection form $\Omega=\left[  \left(  i\partial_{\tau
}t\right)  t^{-1}\right]  _{SU\left(  2,2\right)  }$)$.$ Therefore, the kappa
transformations (\ref{kappa}) inserted in (\ref{actionV3}) give
\begin{equation}
\delta_{\kappa}S_{h}=\int d\tau\left\{  -\frac{1}{2}\left(  \delta_{\kappa
}A^{ij}\right)  X_{i}\cdot X_{j}+iTr\left(  \left(  D_{\tau}t\right)
t^{-1}\left(
\genfrac{}{}{0pt}{}{0}{-\bar{K}\mathcal{L}}%
\genfrac{}{}{0pt}{}{\mathcal{L}K}{0}%
\right)  \right)  \right\}  . \label{delS}%
\end{equation}
In computing the second term the derivative terms that contain $\partial
_{\tau}K$ have dropped out in the trace. Using Eq.(\ref{kappaform}) we see
that
\begin{align}
\mathcal{L}K  &  =\frac{1}{4i}\left(  \varepsilon^{li}X_{l}^{M}X_{i}%
^{N}\right)  X_{j}^{L}\Gamma_{MN}\Gamma_{L}\kappa^{j}\\
&  =\frac{1}{4i}\varepsilon^{li}X_{l}^{M}X_{i}^{N}X_{j}^{L}\left(
\Gamma_{MNL}+\eta_{NL}\Gamma_{M}-\eta_{ML}\Gamma_{N}\right)  \kappa^{l}\\
&  =\frac{1}{2i}\varepsilon^{li}X_{i}\cdot X_{j}\left(  X_{l}\cdot\Gamma
\kappa^{j}\right)
\end{align}
The completely antisymmetric $X_{i}^{M}X_{j}^{N}X_{l}^{L}\Gamma_{MNL}$ term in
the second line vanishes since $i,j,l$ can only take two values. The crucial
observation is that the remaining term in $\mathcal{L}K$ is proportional to
the dot products $X_{i}\cdot X_{j}.$ Therefore the second term in (\ref{delS})
is cancelled by the first term by choosing the appropriate $\delta_{\kappa
}A^{ij}$ in Eq.(\ref{delS}), thus establishing the kappa symmetry.

The local kappa transformations (\ref{kappa}) are also a symmetry of the
global SU$\left(  2,3\right)  _{R}$ charges $\delta_{\kappa}\mathcal{J}%
=\delta_{\kappa}J_{0}=\delta_{\kappa}j_{A}=0$ provided the constraints
$X_{i}\cdot X_{j}=0$ are used. Hence these charges are kappa invariant in the
physical sector.

We have established the global SO$\left(  4,2\right)  $ and local Sp$\left(
2,R\right)  \times($3/4 kappa)$\times$U$\left(  1\right)  $ symmetries of the
phase space action (\ref{actionV2}) in 4+2 dimensions. From it we can derive
all of the phase space actions of the systems listed in section (\ref{s1}) by
making various gauge choices for the local Sp$\left(  2,R\right)  \times($3/4
kappa)$\times$U$\left(  1\right)  $ symmetries. This was demonstrated for the
spinless case $h=0$ in \cite{twistorBP1}. The gauge choices for $X^{M},P^{M}$
discussed in \cite{twistorBP1} now need to be supplemented with gauge choices
for $V_{A},\bar{V}^{A}$ by using the kappa$\times$U$\left(  1\right)  $ local symmetries.

Here we demonstrate the gauge fixing described above for the massless particle
of any spin $h.$ The kappa symmetry effectively has 3 complex gauge parameters
as explained in footnote (\ref{effective}). If the kappa gauge is fixed by
using two of its parameters we reach the following forms
\begin{equation}
V_{A}\rightarrow\left(
\genfrac{}{}{0pt}{}{v^{\dot{\alpha}}}{0}%
\right)  ,\;\bar{V}^{A}\rightarrow\left(  0\;\bar{v}^{\alpha}\right)
,\;\;\bar{V}V\rightarrow0,\;\left(  1-2hV\bar{V}\right)  ^{-1/2}%
\rightarrow\left(
\genfrac{}{}{0pt}{}{1~}{0}%
\genfrac{}{}{0pt}{}{~~hv\bar{v}}{1}%
\right)  . \label{Vfixed}%
\end{equation}
By inserting this gauge fixed form of $V,$ and the gauge fixed form of $X,P$
given in Eq.(\ref{massless}), into the action (\ref{actionV2}) we immediately
recover the SL$\left(  2,C\right)  $ covariant action of Eq.(\ref{actionps}).
The U$\left(  1\right)  $ gauge symmetry is intact. The kappa symmetry of the
action of Eq.(\ref{actionps}) discussed in Eqs.(\ref{kappa1},\ref{kappa2}) is
the residual 1/4 kappa symmetry of the more general action ((\ref{actionV2}).

For other examples of gauge fixing that generates some of the systems in the
list of section (\ref{s1}) see our related paper \cite{twistorBO}.

\section{General Twistor Transform (Classical)}

The various formulations of spinning particles described above all contain
gauge degrees of freedom of various kinds. However, they all have the global
symmetry SU$\left(  2,2\right)  $=SO$\left(  4,2\right)  $ whose conserved
charges $\mathcal{J}_{A}^{~B}$ are gauge invariant in all the formulations.
The most symmetric 2T-physics version gave the $\mathcal{J}_{A}^{~B}$ as
embedded in SU$\left(  2,3\right)  _{R}$ in the SU$\left(  2,2\right)  $
projected form in Eq.(\ref{conserved})
\begin{equation}
\mathcal{J=}\left[  g^{-1}\left(
\genfrac{}{}{0pt}{}{\mathcal{L}}{0}%
\genfrac{}{}{0pt}{}{0}{0}%
\right)  g\right]  _{SU\left(  2,2\right)  }.
\end{equation}
Since this is gauge invariant, when gauge fixed, it must agree with the
Noether charges computed in any version of the theory. So we can equate the
general phase space version of Eq.(\ref{J}) with the twistor version that
follows from the Noether currents of (\ref{action}) as follows
\begin{equation}
\mathcal{J}=Z^{\left(  h\right)  }\bar{Z}^{\left(  h\right)  }-\frac{1}%
{4}Tr\left(  Z^{\left(  h\right)  }\bar{Z}^{\left(  h\right)  }\right)
=\frac{1}{\sqrt{1-2hV\bar{V}}}\mathcal{L}\frac{1}{\sqrt{1-2hV\bar{V}}}%
-\frac{1}{4}J_{0} \label{JZV}%
\end{equation}
The trace corresponds to the U$\left(  1\right)  $ charge $J_{0}=Tr\left(
Z^{\left(  h\right)  }\bar{Z}^{\left(  h\right)  }\right)  ,$ so%
\begin{equation}
\mathcal{J+}\frac{1}{4}J_{0}=Z^{\left(  h\right)  }\bar{Z}^{\left(  h\right)
}=\frac{1}{\sqrt{1-2hV\bar{V}}}\mathcal{L}\frac{1}{\sqrt{1-2hV\bar{V}}}.
\label{equality}%
\end{equation}
In the case of $h=0$ this becomes
\begin{equation}
Z^{\left(  0\right)  }\bar{Z}^{\left(  0\right)  }=\mathcal{L}.
\label{zerotwistor}%
\end{equation}
Therefore the equality (\ref{equality}) is solved up to an irrelevant phase by%
\begin{equation}
Z^{\left(  h\right)  }=\frac{1}{\sqrt{1-2hV\bar{V}}}Z^{\left(  0\right)  }.
\label{general}%
\end{equation}
By inserting (\ref{zerotwistor}) into the constraint (\ref{constraintVL}) we
learn a new form of the constraints%
\begin{equation}
\bar{V}Z^{\left(  0\right)  }=\sqrt{1-2h\bar{V}V},\;\;\bar{V}Z^{\left(
h\right)  }=1. \label{constr2}%
\end{equation}
In turn, this implies%
\begin{equation}
Z^{\left(  0\right)  }=\frac{\mathcal{L}V}{\sqrt{1-2h\bar{V}V}} \label{z0LV}%
\end{equation}
which is consistent\footnote{To see this, we note that Eqs.(\ref{zerotwistor}%
,\ref{constr2}) lead to $\frac{\mathcal{L}V\bar{V}\mathcal{L}}{1-2h\bar{V}%
V}=\frac{Z^{\left(  0\right)  }\bar{Z}^{\left(  0\right)  }V\bar{V}Z^{\left(
0\right)  }\bar{Z}^{\left(  0\right)  }}{1-2h\bar{V}V}=Z^{\left(  0\right)
}\bar{Z}^{\left(  0\right)  }=\mathcal{L}.$} with $Z^{\left(  0\right)  }%
\bar{Z}^{\left(  0\right)  }=\mathcal{L}$ , and its vanishing trace $\bar
{Z}^{\left(  0\right)  }Z^{\left(  0\right)  }=0$ since $\mathcal{LL}=0$ (due
to $X^{2}=P^{2}=X\cdot P=0$). Putting it all together we then have%
\begin{equation}
Z^{\left(  h\right)  }=\frac{1}{\sqrt{1-2hV\bar{V}}}\mathcal{L}V\frac{1}%
{\sqrt{1-2h\bar{V}V}}=\left(  \mathcal{J+}\frac{1}{4}J_{0}\right)  V.
\label{general2}%
\end{equation}
We note that this $Z^{\left(  h\right)  }$ is proportional to the
non-conserved coset part of the SU$\left(  2,3\right)  $ charges $J_{2,3}$,
that is $j_{A}=\sqrt{J_{0}}Z^{\left(  h\right)  }$ given in
Eqs.(\ref{conserved},\ref{rho}) or (\ref{J0}), when $g$ and $\mathcal{L}$ are
replaced by their gauge fixed forms, and use the constraint\footnote{For the
high spin version ($\tilde{A}=0$) we don't use the constraint. Instead, we use
$Z^{\left(  h\right)  }=\frac{1}{\sqrt{1-2hV\bar{V}}}Z^{\left(  0\right)  }$
only in its form (\ref{general}), and note that, after using
Eq.(\ref{zerotwistor}), the $j_{A}$ in Eq.(\ref{J0}) takes the form
$j_{A}=\sqrt{J_{0}}Z^{\left(  h\right)  }$ with$\sqrt{J_{0}}=\frac{\bar{Z}%
^{0}V\sqrt{2h}}{\sqrt{1-2h\bar{V}V}},$ and it is possible to rescale $h$ away
everywhere $\sqrt{2h}V\rightarrow V.$} $J_{0}=2h$.

The key for the general twistor transform for any spin is Eq.(\ref{general}),
or equivalently (\ref{general2}). The general twistor transform between
$Z^{\left(  0\right)  }$ and $X^{M},P^{M}$ which satisfies $Z^{\left(
0\right)  }\bar{Z}^{\left(  0\right)  }=\mathcal{L}$ is already given in
\cite{twistorBP1} as
\begin{equation}
Z^{\left(  0\right)  }=\left(
\begin{array}
[c]{c}%
\mu^{\left(  0\right)  }\\
\lambda^{\left(  0\right)  }%
\end{array}
\right)  ,\;\;\left(  \mu^{\left(  0\right)  }\right)  ^{\dot{\alpha}}%
=-i\frac{X^{\mu}}{X^{+^{\prime}}}\left(  \bar{\sigma}_{\mu}\lambda^{\left(
0\right)  }\right)  ^{\dot{\alpha}},\;\;\lambda_{\alpha}^{\left(  0\right)
}\bar{\lambda}_{\dot{\beta}}^{\left(  0\right)  }=\left(  X^{+}P^{\mu}-X^{\mu
}P^{+}\right)  \left(  \sigma_{\mu}\right)  _{\alpha\dot{\beta}}. \label{z0}%
\end{equation}
Note that $\left(  X^{+}P^{\mu}-X^{\mu}P^{+}\right)  $ is compatible with the
requirement that any SL$\left(  2,C\right)  $ vector constructed as
$\lambda_{\alpha}^{\left(  0\right)  }\bar{\lambda}_{\dot{\beta}}^{\left(
0\right)  }$ must be lightlike. This property is satisfied thanks to the
Sp$\left(  2,R\right)  $ constraints $X^{2}=P^{2}=X\cdot P=0$ in 4+2
dimensions, thus allowing a particle of \textit{any mass} in the $3+1$
subspace (since $P^{\mu}P_{\mu}$ is not restricted to be lightlike). Besides
satisfying $Z^{\left(  0\right)  }\bar{Z}^{\left(  0\right)  }=\mathcal{L}$,
this $Z^{\left(  0\right)  }$ also satisfies $\bar{Z}^{\left(  0\right)
}Z^{\left(  0\right)  }=0,$ as well as the canonical properties of twistors.
Namely, $Z^{\left(  0\right)  }$ has the property \cite{twistorBP1}%
\begin{equation}
\int d\tau~\bar{Z}^{\left(  0\right)  }\partial_{\tau}Z^{\left(  0\right)
}=\int d\tau~\dot{X}^{M}P_{M}. \label{canonical0}%
\end{equation}
From here, by gauge fixing the Sp$\left(  2,R\right)  $ gauge symmetry, we
obtain the twistor transforms for all the systems listed in section (\ref{s1})
for $h=0$ directly from Eq.(\ref{z0}), as demonstrated in \cite{twistorBP1}.
All of that is now generalized at once to any spin $h$ through
Eq.(\ref{general}). Hence (\ref{general}) together with (\ref{z0}) tell us how
to construct explicitly the \textit{general} twistor $Z_{A}^{\left(  h\right)
}$ in terms of spin \& phase space degrees of freedom $X^{M},P^{M},V_{A}%
,\bar{V}^{A}.$ Then the Sp$\left(  2,R\right)  $ and kappa gauge symmetries
that act on $X^{M},P^{M},V_{A},\bar{V}^{A}$ can be gauge fixed for any spin
$h,$ to give the \textit{specific} twistor transform for any of the systems
under consideration.

We have already seen in Eq.(\ref{JZV}) that the twistor transform
(\ref{general}) relates the conserved SU$\left(  2,2\right)  $ charges in
twistor and phase space versions. Let us now verify that (\ref{general})
provides the transformation between the twistor action (\ref{action}) and the
spin \& phase space action (\ref{actionV2}). We compute the canonical
structure as follows%
\begin{align}
\int d\tau~\bar{Z}^{\left(  h\right)  }\partial_{\tau}Z^{\left(  h\right)  }
&  =\int d\tau~\bar{Z}^{\left(  0\right)  }\frac{1}{\sqrt{1-2hV\bar{V}}%
}\partial_{\tau}\left(  \frac{1}{\sqrt{1-2hV\bar{V}}}Z^{\left(  0\right)
}\right) \\
&  =\int d\tau~\left\{
\begin{array}
[c]{c}%
\bar{Z}^{\left(  0\right)  }\frac{1}{\sqrt{1-2hV\bar{V}}}\left(
\partial_{\tau}\frac{1}{\sqrt{1-2hV\bar{V}}}\right)  Z^{\left(  0\right)  }\\
+\bar{Z}^{\left(  0\right)  }\frac{1}{1-2hV\bar{V}}\partial_{\tau}Z^{\left(
0\right)  }%
\end{array}
\right\} \\
&  =\int d\tau~\left\{  \dot{X}\cdot P+Tr\left[  \left(  i\partial_{\tau
}t\right)  t^{-1}\left(
\genfrac{}{}{0pt}{}{\mathcal{L}}{0}%
\genfrac{}{}{0pt}{}{0}{0}%
\right)  \right]  \right\}
\end{align}
The last form is the canonical structure of spin \& phase space as given in
(\ref{actionV2}). To prove this result we used Eq.(\ref{canonical0}), footnote
(\ref{powers}), and the other properties of $Z^{\left(  0\right)  }$ including
Eqs.(\ref{zerotwistor}-\ref{z0LV}), as well as the constraints $X^{2}%
=P^{2}=X\cdot P=0,$ and dropped some total derivatives. This proves that the
canonical properties of $Z^{\left(  h\right)  }$ determine the canonical
properties of spin \& phase space degrees of freedom and vice versa.

Then, including the terms that impose the constraints, the twistor action
(\ref{action}) and the phase space action (\ref{actionV2}) are equivalent. Of
course, this is expected since they are both gauge fixed versions of the
master action (\ref{actionV}), but is useful to establish it also directly via
the general twistor transform given in Eq.(\ref{general}).

\section{Quantum Master Equation, Spectrum, and Dualities}

In this section we derive the quantum algebra of the gauge invariant
observables $\mathcal{J}_{A}^{~B}$ and $J_{0}$ which are the conserved charges
of [SU$\left(  2,2\right)  \times$U$\left(  1\right)  ]_{R}.$ Since these are
gauge invariant symmetry currents they govern the system in any of its gauge
fixed versions, including in any of its versions listed in section (\ref{s1}).
From the quantum algebra we deduce the constraints among the physical
observables $\mathcal{J}_{A}^{~B}$,$J_{0}$ and quantize the theory
covariantly. Among other things, we compute the Casimir eigenvalues of the
unitary irreducible representation of SU$\left(  2,2\right)  $ which
classifies the physical states in any of the gauge fixed version of the theory
(with the different 1T-physics interpretations listed in section (\ref{s1})).

The simplest way to quantize the theory is to use the twistor variables, and
from them compute the gauge invariant properties that apply in any gauge fixed
version. We will apply the covariant quantization approach, which means that
the constraint due to the U$\left(  1\right)  $ gauge symmetry will be applied
on states. Since the quantum variables will generally not satisfy the
constraints, we will call the quantum twistors in this section $Z_{A},\bar
{Z}^{A}$ to distinguish them from the classical $Z_{A}^{\left(  h\right)
},\bar{Z}^{\left(  h\right)  A}$ of the previous sections that were
constrained at the classical level. So the formalism in this section can also
be applied to the high spin theories (discussed in several footnotes up to
this point in the paper) by ignoring the constraint on the states.

According to the twistor action (\ref{action}) $Z_{A}$ and $i\bar{Z}^{A}$ (or
equivalently $\lambda_{\alpha}$ and $i\bar{\mu}^{\alpha}$) are canonical
conjugates. Therefore the quantum rules (equivalent to spin \& phase space
quantum rules) are
\begin{equation}
\left[  Z_{A},\bar{Z}^{B}\right]  =\delta_{A}^{~B}. \label{ZZ1}%
\end{equation}
These quantum rules, as well as the action, are manifestly invariant under
SU$\left(  2,2\right)  .$ In covariant SU$\left(  2,2\right)  $ quantization
the Hilbert space contains states which do not obey the U$\left(  1\right)  $
constraint on the twistors. At the classical level the constraint was
$J_{0}=\bar{Z}Z=2h,$ but in covariant quantization this is obeyed only by the
U$\left(  1\right)  $ gauge invariant subspace of the Hilbert space which we
call the physical states. The quantum version of the constraint requires
$\hat{J}_{0}$ as a Hermitian operator applied on states (we write it as
$\hat{J}_{0}$ to distinguish it from the classical version)
\begin{equation}
\hat{J}_{0}=\frac{1}{2}\left(  Z_{A}\bar{Z}^{A}+\bar{Z}^{A}Z_{A}\right)
,\;\;\hat{J}_{0}|phys\rangle=2h|phys\rangle. \label{hhat}%
\end{equation}
The operator $\hat{J}_{0}$ has non-trivial commutation relations with
$Z_{A},\bar{Z}^{~A}$ which follow from the basic commutation rules above%
\begin{equation}
\left[  \hat{J}_{0},Z_{A}\right]  =-Z_{A},\;\;\;\left[  \hat{J}_{0},\bar
{Z}^{~A}\right]  =\bar{Z}^{~A}. \label{[hZ]}%
\end{equation}
By rearranging the orders of the quantum operators $Z_{A}\bar{Z}^{A}=\bar
{Z}^{A}Z_{A}+4$ we can extract from (\ref{hhat}) the following relations%
\begin{equation}
\bar{Z}Z=\hat{J}_{0}-2,\;\;Tr\left(  Z\bar{Z}\right)  =\hat{J}_{0}+2.
\label{ZZh}%
\end{equation}
Furthermore, by using Noether's theorem for the twistor action (\ref{action})
we can derive the 15 generators of SU$\left(  2,2\right)  $ in terms of the
twistors and write them as a traceless $4\times4$ matrix $\mathcal{J}_{A}%
^{~B}$ at the quantum level as follows
\begin{equation}
\mathcal{J}_{A}^{~B}=Z_{A}\bar{Z}^{B}-\frac{1}{4}Tr\left(  Z\bar{Z}\right)
\delta_{A}^{~B}=\left(  Z\bar{Z}-\frac{\hat{J}_{0}+2}{4}\right)  _{A}^{~B}.
\label{JZZ}%
\end{equation}
In this expression the order of the quantum operators matters and gives rise
to the shift $J_{0}\rightarrow\hat{J}_{0}+2$ in contrast to the corresponding
classical expression. The commutation rules among the generators
$\mathcal{J}_{A}^{~B}$ and the $Z_{A},\bar{Z}^{A}$ are computed from the basic
commutators (\ref{ZZ1}),
\begin{align}
\left[  \mathcal{J}_{A}^{~B},Z_{C}\right]   &  =-\delta_{C}^{~B}Z_{A}+\frac
{1}{4}Z_{C}~\delta_{A}^{~B},\;\;\left[  \mathcal{J}_{A}^{~B},\bar{Z}%
^{D}\right]  =\delta_{A}^{~D}\bar{Z}^{B}-\frac{1}{4}\bar{Z}^{D}~\delta
_{A}^{~B}\label{commsJZ}\\
\left[  \mathcal{J}_{A}^{~B},\mathcal{J}_{C}^{~D}\right]   &  =\delta_{A}%
^{~D}\mathcal{J}_{C}^{~B}-\delta_{C}^{~B}\mathcal{J}_{A}^{~D},\;\;\left[
\hat{J}_{0},\mathcal{J}_{A}^{~B}\right]  =0. \label{commsJ1}%
\end{align}
We see from these that the gauge invariant observables $\mathcal{J}_{A}^{~B}$
satisfy the SU$\left(  2,2\right)  $ Lie algebra, while the $Z_{A},\bar{Z}%
^{A}$ transform like the quartets $4,\bar{4}$ of SU$\left(  2,2\right)  .$
Note that the operator $\hat{J}_{0}$ commutes with the generators
$\mathcal{J}_{A}^{~B},$ therefore $\mathcal{J}_{A}^{~B}$ is U$\left(
1\right)  $ gauge invariant, and furthermore $\hat{J}_{0}$ must be a function
of the Casimir \textit{operators} of SU$\left(  2,2\right)  .$ When $\hat
{J}_{0}$ takes the value $2h$ on physical states, then the Casimir operators
also will have eigenvalues on physical states which determine the SU$\left(
2,2\right)  $ representation in the physical sector.

From the quantum rules (\ref{[hZ]}), it is evident that the U$\left(
1\right)  $ generator $\hat{J}_{0}$ can only have integer eigenvalues since it
acts like a number of operator. More directly, through Eq.(\ref{ZZh}) it is
related to the number operator $\bar{Z}Z$. Therefore the theory is consistent
at the quantum level (\ref{hhat}) provided $2h$ is an integer.

Let us now compute the square of the matrix $\mathcal{J}_{A}^{~B}.$ By using
the form (\ref{JZZ}) we have $\left(  \mathcal{JJ}\right)  $ = $\left(
Z\bar{Z}-\frac{\hat{J}_{0}+2}{4}\right)  \left(  Z\bar{Z}-\frac{\hat{J}_{0}%
+2}{4}\right)  $ = $Z\bar{Z}Z\bar{Z}-2\frac{\hat{J}_{0}+2}{4}Z\bar{Z}+\left(
\frac{\hat{J}_{0}+2}{4}\right)  ^{2}$ where we have used $\left[  \hat{J}%
_{0},Z_{A}\bar{Z}^{B}\right]  =0.$ Now we elaborate $\left(  Z\bar{Z}Z\bar
{Z}\right)  _{A}^{B}$ = $Z_{A}\left(  \hat{J}_{0}-2\right)  \bar{Z}^{B}$ =
$\left(  \hat{J}_{0}-1\right)  Z_{A}\bar{Z}^{B}$ where we first used
(\ref{ZZh}) and then (\ref{[hZ]}). Finally we note from (\ref{JZZ}) that
$Z_{A}\bar{Z}^{B}$ = $\mathcal{J}_{A}^{~B}+\frac{\hat{J}_{0}+2}{4}\delta
_{A}^{~B}$. Putting these observations together we can rewrite the right hand
side of $\left(  \mathcal{JJ}\right)  $ in terms of $\mathcal{J}$ and $\hat
{J}_{0}$ as follows\footnote{A similar structure at the classical level can be
easily computed by squaring the expression for $\mathcal{J}$ in Eq.(\ref{JZV})
and applying the classical constraint $J_{0}=\bar{Z}^{A}Z_{A}=2h.$ This yields
the classical version $\mathcal{J}_{A}^{~C}\mathcal{J}_{C}^{~B}=\frac{J_{0}%
}{2}\mathcal{J}_{A}^{~B}+\frac{3}{16}J_{0}^{2}\delta_{A}^{~B}=h\mathcal{J}%
_{A}^{~B}+\frac{3}{4}h^{2}\delta_{A}^{~B},$ which is different than the
quantum equation (\ref{JJtoJ2}). Thus, the quadratic Casimir at the
\textit{classical} level is computed as $C_{2}=\frac{3}{4}J_{0}^{2}=3h^{2}$
which is different than the quantum value in (\ref{twistingCasimirsa}).
\label{classJ}}
\begin{equation}
\left(  \mathcal{JJ}\right)  =\left(  \frac{\hat{J}_{0}}{2}-2\right)
\mathcal{J+}\frac{3}{16}\left(  \hat{J}_{0}^{2}-4\right)  . \label{JJtoJ2}%
\end{equation}
This equation is a constraint satisfied by the global [SU$\left(  2,2\right)
\times$U$\left(  1\right)  ]_{R}$ charges $\mathcal{J}_{A}^{~B},\hat{J}_{0}$
which are gauge invariant physical observables. It is a correct equation for
all the states in the theory, including those that do not satisfy the
U$\left(  1\right)  $ constraint (\ref{hhat}). We call this the
\textit{quantum master equation} because it will determine completely all the
SU$\left(  2,2\right)  $ properties of the physical states for all the systems
listed in section (\ref{s1}) for any spin.

By multiplying the master equation with $\mathcal{J}$ and using (\ref{JJtoJ2})
again we can compute~$\mathcal{JJJ}$. Using this process repeatedly we find
all the powers of the matrix $\mathcal{J}$
\begin{equation}
\left(  \mathcal{J}\right)  ^{n}=\alpha_{n}\mathcal{J}+\beta_{n}, \label{n}%
\end{equation}
where%
\begin{align}
\alpha_{n}(\hat{J}_{0})  &  =\frac{1}{\hat{J}_{0}-1}\left[  \left(  \frac
{3}{4}\left(  \hat{J}_{0}-2\right)  \right)  ^{n}-\left(  \frac{-1}{4}\left(
\hat{J}_{0}+2\right)  \right)  ^{n}\right]  ,\;\;\\
\beta_{n}(\hat{J}_{0})  &  =\frac{3}{16}\left(  \hat{J}_{0}^{2}-4\right)
~\alpha_{n-1}(\hat{J}_{0}).
\end{align}
Remarkably, these formulae apply to all powers, including negative powers of
the matrix $\mathcal{J}.$ Using this result, any function of the matrix
$\mathcal{J}$ constructed as a Taylor series takes the form
\begin{equation}
f\left(  \mathcal{J}\right)  =\alpha\left(  \hat{J}_{0}\right)  \mathcal{J+}%
\beta\left(  \hat{J}_{0}\right)  \label{f}%
\end{equation}
where%
\begin{align}
\alpha\left(  \hat{J}_{0}\right)   &  =\frac{1}{\hat{J}_{0}-1}\left[  f\left(
\frac{3}{4}\left(  \hat{J}_{0}-2\right)  \right)  -f\left(  \frac{-1}%
{4}\left(  \hat{J}_{0}+2\right)  \right)  \right]  ,\;\\
\beta\left(  \hat{J}_{0}\right)   &  =\frac{1}{\hat{J}_{0}-1}\left[
\frac{\left(  \hat{J}_{0}+2\right)  }{4}f\left(  \frac{3}{4}\left(  \hat
{J}_{0}-2\right)  \right)  +\frac{3\left(  \hat{J}_{0}-2\right)  }{4}f\left(
\frac{-1}{4}\left(  \hat{J}_{0}+2\right)  \right)  \right]  .
\end{align}

We can compute all the Casimir operators by taking the trace of $\mathcal{J}%
^{n}$ in Eq.(\ref{n}), so we find\footnote{Other definitions of $C_{n}$could
differ from ours by normalization or linear combinations of the $Tr\left(
\mathcal{J}^{n}\right)  $.}
\begin{equation}
C_{n}(\hat{J}_{0})\equiv Tr\left(  \mathcal{J}\right)  ^{n}=4\beta_{n}(\hat
{J}_{0})=\frac{3}{4}\left(  \hat{J}_{0}^{2}-4\right)  ~\alpha_{n-1}(\hat
{J}_{0}).
\end{equation}
In particular the quadratic, cubic and quartic Casimir operators of SU$\left(
2,2\right)  =$SO$\left(  6,2\right)  $ are computed at the quantum level as
\begin{align}
C_{2}(\hat{J}_{0})  &  =\frac{3}{4}\left(  \hat{J}_{0}^{2}-4\right)
,\;\;\;C_{3}(\hat{J}_{0})=\frac{3}{8}\left(  \hat{J}_{0}^{2}-4\right)  \left(
\hat{J}_{0}-4\right)  ,\;\label{twistingCasimirsa}\\
C_{4}(\hat{J}_{0})  &  =\frac{3}{64}\left(  \hat{J}_{0}^{2}-4\right)  \left(
7\hat{J}_{0}^{2}-32\hat{J}_{0}+52\right)  . \label{twistingCasimirb}%
\end{align}
The eigenvalue of the operator $\hat{J}_{0}$ on physical states $\hat{J}%
_{0}|phys\rangle=2h|phys\rangle$ completely fixes the unitary SU$\left(
2,2\right)  $ representation that classifies the physical states, since the
most general representation of SO$\left(  4,2\right)  $ is labeled by the
three independent eigenvalues of $C_{2},C_{3}$ and $C_{4}.$ Obviously, this
result is a special representation of SU$\left(  2,2\right)  $ since all the
Casimir eigenvalues are determined in terms of a single half integer number
$h.$ Therefore we conclude that\emph{ }all of the systems listed in section
(\ref{s1}) share the very same unitary representation of SU$\left(
2,2\right)  $ with the same Casimir eigenvalues given above.

In particular, for spinless particles $\left(  \hat{J}_{0}\rightarrow
h=0\right)  $ we obtain $C_{2}=-3$, $C_{3}=6$, $C_{4}=-\frac{39}{4},$ which is
the unitary \textit{singleton} representation of SO$\left(  4,2\right)
=$SU$\left(  2,2\right)  $. This is in agreement with previous covariant
quantization of the spinless particle in any dimension directly in phase space
in $d+2$ dimensions, which gave for the SO$\left(  d,2\right)  $ Casimir the
eigenvalue as $C_{2}=\frac{1}{2}L_{MN}L^{MN}\rightarrow1-d^{2}/4$ on physical
states that satisfy $X^{2}=P^{2}=X\cdot P=0$ \cite{2treviews}. So, for $d=4$
we get $C_{2}=-3$ in agreement with the quantum twistor computation above.
Note that the classical computation either in phase space or twistor space
would give the wrong answer $C_{2}=0$ when orders of canonical conjugates are
ignored and constraints used classically.

Of course, having the same SU$\left(  2,2\right)  $ Casimir eigenvalue is one
of the \textit{infinite number of duality relations} among these systems that
follow from the more general twistor transform or the master 2T-physics theory
(\ref{actionV}). All dualities of these systems amount to all quantum
functions of the gauge invariants $\mathcal{J}_{A}^{~B}$ that take the same
gauge invariant values in any of the physical Hilbert spaces of the systems
listed in section (\ref{s1}).

All the physical information on the relations among the physical observables
is already captured by the quantum master equation (\ref{JJtoJ2}), so it is
sufficient to concentrate on it. The predicted duality, including these
relations, can be tested at the quantum level by computing and verifying the
equality of an infinite number of matrix elements of the master equation
between the dually related quantum states for the systems listed in section
(\ref{s1}). In the case of the Casimir operators $C_{n}$ the details of the
individual states within a representation is not relevant, so that computation
whose result is given above is among the simplest computations that can be
performed on the systems listed in section (\ref{s1}) to test our duality
predictions. This test was performed successfully for $h=0$ at the quantum
level for some of these systems directly in their own phase spaces
\cite{2tHandAdS}, verifying for example, that the free massless particle, the
hydrogen atom, the harmonic oscillator, the particle on AdS spaces, all have
the same Casimir eigenvalues $C_{2}=-3$, $C_{3}=6$, $C_{4}=-\frac{39}{4}$ at
the quantum level.

Much more elaborate tests of the dualities can be performed both at the
classical and quantum levels by computing any function of the gauge invariant
$\mathcal{J}_{A}^{~B}$ and checking that it has the same value when computed
in terms of the spin \& phase space of any of the systems listed in section
(\ref{s1}). At the quantum level all of these systems have the same Casimir
eigenvalues of the $C_{n}$ for a given $h.$ So their spectra must correspond
to the same unitary irreducible representation of SU$\left(  2,2\right)  $ as
seen above. But the rest of the labels of the representation correspond to
simultaneously commuting operators that include the Hamiltonian. The
Hamiltonian of each system is some operator constructed from the observables
$\mathcal{J}_{A}^{~B},$ and so are the other simultaneously diagonalizable
observables. Therefore, the different systems are related to one another by
unitary transformations that sends one Hamiltonian to another, but staying
within the same representation. These unitary transformations are the quantum
versions of the gauge transformations of Eq.(\ref{locals}), and so they are
the duality transformations at the quantum level. In particular the twistor
transform applied to any of the systems is one of those duality
transformations. By applying the twistor transforms we can map the Hilbert
space of one system to another, and then compute any function of the gauge
invariant $\mathcal{J}_{A}^{~B}$ between dually related states of different
systems. The prediction is that all such computations within different systems
must give the same result.

Given that $\mathcal{J}_{A}^{~B}$ is expressed in terms of rather different
phase space and spin degrees of freedom in each dynamical system with a
different Hamiltonian, this predicted duality is remarkable. 1T-physics simply
is not equipped to explain why or for which systems there are such dualities,
although it can be used to check it. The origin as well as the proof of the
duality is the unification of the systems in the form of the 2T-physics master
action of Eq.(\ref{actionV}) in 4+2 dimensions. The existence of the
dualities, which can laboriously be checked using 1T-physics, is the evidence
that the underlying spacetime is more beneficially understood as being a
spacetime in 4+2 dimensions.

\section{Quantum Twistor Transform}

We have established a master equation for physical observables $\mathcal{J}$
at the quantum level. Now, we also want to establish the twistor transform at
the quantum level expressed as much as possible in terms of the gauge
invariant physical quantum observables $\mathcal{J}$. To this end we write the
master equation (\ref{JJtoJ2}) in the form%
\begin{equation}
\left(  \mathcal{J}-\frac{3}{4}\left(  \hat{J}_{0}-2\right)  \right)  \left(
\mathcal{J}+\frac{1}{4}\left(  \hat{J}_{0}+2\right)  \right)  =0.
\label{master}%
\end{equation}
Recall the quantum equation (\ref{JZZ}) $\mathcal{J}+\frac{\hat{J}_{0}+2}%
{4}=Z\bar{Z},$ so the equation above is equivalent to%
\begin{equation}
\left(  \mathcal{J}-\frac{3}{4}\left(  \hat{J}_{0}-2\right)  \right)  Z=0.
\label{eigenvalue}%
\end{equation}
This is a $4\times4$ matrix eigenvalue equation with operator entries. The
general solution is
\begin{equation}
Z=\left(  \mathcal{J}+\frac{1}{4}\left(  \hat{J}_{0}+2\right)  \right)
\hat{V} \label{solution}%
\end{equation}
where $\hat{V}_{A}$ is any spinor up to a normalization. This is verified by
using the master equation (\ref{master}) which gives $\left(  \mathcal{J}%
-\frac{3}{4}\left(  \hat{J}_{0}-2\right)  \right)  Z=\left(  \mathcal{J}%
-\frac{3}{4}\left(  \hat{J}_{0}-2\right)  \right)  \left(  \mathcal{J}%
+\frac{1}{4}\left(  \hat{J}_{0}+2\right)  \right)  \hat{V}=0.$ Noting that the
solution (\ref{solution}) has the same form as the classical version of the
twistor transform in Eq.(\ref{general2}), except for the quantum shift
$J_{0}\rightarrow\hat{J}_{0}+2$, we conclude that the $\hat{V}_{A}$ introduced
above is the quantum version of the $V_{A}$ discussed earlier (up to a
possible renormalization\footnote{The quantum version of $\hat{V}$ is valid in
the whole Hilbert space, not only in the subspace that satisfies the U$\left(
1\right)  $ constraint $\hat{J}_{0}\rightarrow2h.$ In particular, in the high
spin version, already at the classical level we must take $\hat{V}=V(\sqrt
{2h}/\sqrt{J_{0}})$ and then rescale it $V\sqrt{2h}\rightarrow V$ as described
in previous footnotes. So in the full quantum Hilbert space we must take
$\hat{V}=\sqrt{2h}V(\hat{J}_{0}+\gamma)^{-1/2}$ (or the rescaled version
$V\sqrt{2h}\rightarrow V$) with the possibly quantum shifted operator
$(\hat{J}_{0}+\gamma)^{-1/2}$.}), as belonging to the coset SU$\left(
2,3\right)  /$[SU$\left(  2,2\right)  \times$U$\left(  1\right)  $].

Now $\hat{V}_{A}$ is a quantum operator whose commutation rules must be
compatible with those of $Z_{A},\bar{Z}^{A},\hat{J}_{0}$ and $\mathcal{J}%
_{A}^{~B}.$ Its commutation rules with $\mathcal{J}_{A}^{~B},\hat{J}_{0}$ are
straightforward and fixed uniquely by the SU$\left(  2,2\right)  \times
$U$\left(  1\right)  $ covariance
\begin{align}
\left[  \hat{J}_{0},\hat{V}_{A}\right]   &  =-\hat{V}_{A},\;\;\left[  \hat
{J}_{0},\overline{\hat{V}}^{A}\right]  =\overline{\hat{V}}^{A},\label{J0V}\\
\left[  \mathcal{J}_{A}^{~B},\hat{V}_{C}\right]   &  =-\delta_{C}^{~B}\hat
{V}_{A}+\frac{1}{4}\hat{V}_{C}~\delta_{A}^{~B}\;,\;~\left[  \mathcal{J}%
_{A}^{~B},\overline{\hat{V}}^{D}\right]  =\delta_{A}^{~D}\overline{\hat{V}%
}^{B}-\frac{1}{4}\overline{\hat{V}}^{D}\delta_{A}^{~B}. \label{JV}%
\end{align}
Other quantum properties of $\hat{V}_{A}$ follow from imposing the quantum
property $\bar{Z}Z=\hat{J}_{0}-2$ in (\ref{ZZh}). Inserting $Z$ of the form
(\ref{solution}), using the master equation, and observing the commutation
rules (\ref{J0V}), we obtain
\begin{equation}
\overline{\hat{V}}\left(  \mathcal{J}+\frac{\hat{J}_{0}+2}{4}\right)  \hat
{V}=1. \label{VJV}%
\end{equation}
This is related to (\ref{constraintVL}) if we take (\ref{J}) into account by
including the quantum shift $J_{0}\rightarrow\hat{J}_{0}+2.$ Considering
(\ref{solution}) this equation may also be written as%
\begin{equation}
\overline{\hat{V}}Z=\bar{Z}\hat{V}=1. \label{VZ}%
\end{equation}
Next we impose $\left[  Z_{A},\bar{Z}^{B}\right]  =\delta_{A}^{~B}$ to deduce
the quantum rules for $[\hat{V}_{A},\overline{\hat{V}}^{B}].$ After some
algebra we learn that the most general form compatible with $\left[
Z_{A},\bar{Z}^{B}\right]  =\delta_{A}^{~B}$ is%

\begin{equation}
\left[  \hat{V}_{A},\overline{\hat{V}}^{B}\right]  =-\frac{\overline{\hat{V}%
}\hat{V}}{\hat{J}_{0}-1}\delta_{A}^{~~B}+\left(  M(\mathcal{J}-3\frac{\hat
{J}_{0}-2}{4})+(\mathcal{J}-3\frac{\hat{J}_{0}-2}{4})\bar{M}\right)
_{A}^{~~B}, \label{structure}%
\end{equation}
where $M_{A}^{~B}$ is some complex matrix and $\bar{M}=\left(  \eta
_{2,2}\right)  M^{\dagger}\left(  \eta_{2,2}\right)  ^{-1}.$ The matrix
$M_{A}^{~B}$ could not be determined uniquely because of the 3/4 kappa gauge
freedom in the choice of $\hat{V}_{A}$ itself.

A maximally gauge fixed version of $\hat{V}_{A}$ corresponds to eliminating 3
of its components $\hat{V}_{2,3,4}=0$ by using the 3/4 kappa symmetry, leaving
only $A\equiv\hat{V}_{1}\neq0.$ Then we find $\overline{\hat{V}}^{1,2,4}=0$
and $\bar{V}^{3}=A^{\dagger}.$ Let us analyze the quantum properties of this
gauge in the context of the formalism above. From Eq.(\ref{VJV}) we determine
$A=\left(  \mathcal{J}_{3}^{~1}\right)  ^{-1/2}e^{-i\phi},$ where $\phi$ is a
phase, and then from Eq.(\ref{solution}) we find $Z_{A}$.
\begin{equation}
Z_{A}=\left(  \mathcal{J}_{A}^{~~1}+\frac{\hat{J}_{0}+2}{4}\delta_{A}%
^{~~1}\right)  \left(  \mathcal{J}_{3}^{~1}\right)  ^{-1/2}e^{-i\phi}%
,\;\;\bar{Z}^{A}=e^{i\phi}\left(  \mathcal{J}_{3}^{~1}\right)  ^{-1/2}\left(
\mathcal{J}_{3}^{~~A}+\frac{\hat{J}_{0}+2}{4}\delta_{3}^{~~A}\right)  .
\label{special}%
\end{equation}
We see that, except for the overall phase, $Z_{A}$ is completely determined in
terms of the gauge invariant $\mathcal{J}_{A}^{~B}.$ We use a set of gamma
matrices $\Gamma^{M}$ given in (\cite{twistorBP1},\cite{susy2t}) to write
$\mathcal{J}_{A}^{~B}=\frac{1}{4i}J^{MN}\left(  \Gamma_{MN}\right)  _{A}^{~B}$
as an explicit matrix so that $Z_{A}$ can be written in terms of the 15
SO$\left(  4,2\right)  =$SU$\left(  2,2\right)  $ generators $J^{MN}.$ We find%
\begin{equation}
Z_{A}=\left(
\begin{array}
[c]{c}%
\frac{1}{2}J^{12}+\frac{1}{2i}J^{+-}+\frac{1}{2i}J^{+^{\prime}-^{\prime}%
}+\frac{\hat{J}_{0}+2}{4}\\
\frac{i}{\sqrt{2}}\left(  J^{+1}+iJ^{+2}\right) \\
J^{+^{\prime}+}\\
\frac{i}{\sqrt{2}}\left(  J^{+^{\prime}1}+iJ^{+^{\prime}2}\right)
\end{array}
\right)  \frac{e^{-i\phi}}{\sqrt{J^{+^{\prime}+}}},\; \label{Za}%
\end{equation}
and $\bar{Z}^{A}=\left(  Z^{\dagger}\eta_{2,2}\right)  ^{A}.$ The orders of
the operators here are important. The basis $M=\pm^{\prime},\pm,i$ with
$i=1,2$ corresponds to using the lightcone combinations $X^{\pm^{\prime}%
}=\frac{1}{\sqrt{2}}\left(  X^{0^{\prime}}\pm X^{1^{\prime}}\right)  ,$
$X^{\pm}=\frac{1}{\sqrt{2}}\left(  X^{0}\pm X^{1}\right)  $.

From our setup above, the $Z_{A},\bar{Z}^{A}$ in (\ref{Za}) are guaranteed to
satisfy the twistor commutation rules $\left[  Z_{A},\bar{Z}^{B}\right]
=\delta_{A}^{~B}$ provided we insure that the $\hat{V}_{A},\overline{\hat{V}%
}^{B}$ have the quantum properties given in Eqs.(\ref{J0V},\ref{JV}%
,\ref{structure}). These are satisfied provided we take the following
non-trivial commutation rules for $\phi$%
\begin{equation}
\left[  \phi,\hat{J}_{0}\right]  =i,\;\left[  \phi,J_{12}\right]  =\frac{i}%
{2}\;\Rightarrow\left[  \hat{J}_{0},e^{\pm i\phi}\right]  =\pm e^{\pm i\phi
},\;[J^{12},e^{\pm i\phi}]=\pm\frac{1}{2}e^{\pm i\phi} \label{phii}%
\end{equation}
while all other commutators between $\phi$ and $J^{MN}$ vanish. Then
(\ref{structure}) becomes $[\hat{V}_{A},\overline{\hat{V}}^{B}]=0,$ so
$M_{A}^{~B}$ vanishes in this gauge. Indeed one can check directly that only
by using the Lie algebra for the $J^{MN},\hat{J}_{0}$ and the commutation
rules for $\phi$ in (\ref{phii}), we obtain $\left[  Z_{A},\bar{Z}^{B}\right]
=\delta_{A}^{~B},$ which a remarkable form of the twistor transform at the
quantum level.

The expression (\ref{Za}) for the twistor is not SU$\left(  2,2\right)  $
covariant. Of course, this is because we chose a non-covariant gauge for
$\hat{V}_{A}.$ However, the global symmetry SU$\left(  2,2\right)  $ is still
intact since the correct commutation rules between the twistors and $J^{MN}$
or the $\mathcal{J}_{A}^{~~B}$ as given in (\ref{commsJZ},\ref{commsJ1}) are
built in, and are automatically satisfied. Therefore, despite the lack of
manifest covariance, the expression for $Z_{A}$ in (\ref{Za}) transforms
covariantly as the spinor of SU$\left(  2,2\right)  .$

It is now evident that one has many choices of gauges for $\hat{V}_{A}.$ Once
a gauge is picked the procedure outlined above will automatically produce the
\textit{quantum} twistor transform in that gauge, and it will have the correct
commutation rules and SU$\left(  2,2\right)  $ properties at the quantum
level. For example, in the SL$\left(  2,C\right)  $ covariant gauge of
Eq.(\ref{Vfixed}), the quantum twistor transform in terms of $J^{MN}$ is%
\begin{equation}
\mu^{\dot{\alpha}}=\frac{1}{4i}J_{\mu\nu}\left(  \bar{\sigma}^{\mu\nu}\right)
_{~\dot{\beta}}^{\dot{\alpha}}v^{\dot{\beta}}+\frac{1}{2i}J^{+^{\prime
}-^{\prime}}v^{\dot{\alpha}},\;\;\lambda_{\alpha}=\frac{1}{\sqrt{2}%
}J^{+^{\prime}\mu}\left(  \sigma_{\mu}\right)  _{\alpha\dot{\beta}}%
v^{\dot{\beta}}.
\end{equation}
with the constraint
\begin{equation}
\frac{1}{\sqrt{2}}\bar{v}\sigma_{\mu}vJ^{+^{\prime}\mu}=1.
\end{equation}
This gauge for $\hat{V}_{M}$ covers several of the systems listed in section
(\ref{s1}). The spinless case was discussed at the classical level in
(\cite{twistorBP1}). The quantum properties of this gauge are discussed in
more detail in (\cite{twistorBO}).

The result for $Z_{A}$ in (\ref{Za}) is a quantum twistor transform that
relies \textit{only on the gauge invariants} $\mathcal{J}_{A}^{~B}$ or
equivalently $J^{MN}.$ It generalizes a similar result in \cite{twistorBP1}
that was given at the classical level. In the present case it is quantum and
with spin. All the information on spin is included in the generators
$J^{MN}=L^{MN}+S^{MN}.$ There are other ways of describing spinning particles.
For example, one can start with a 2T-physics action that uses fermions
$\psi^{M}\left(  \tau\right)  $ \cite{spin2t} instead of our bosonic variables
$V_{A}\left(  \tau\right)  .$ Since we only use the gauge invariant $J^{MN},$
our quantum twistor transform (\ref{solution}) applies to all such
descriptions of spinning particles, with an appropriate relation between
$\hat{V}$ and the new spin degrees of freedom. In particular in the gauge
fixed form of $\hat{V}$ that yields (\ref{Za}) there is no need to seek a
relation between $\hat{V}$ and the other spin degrees of freedom. Therefore,
in the form (\ref{Za}), if the $J^{MN}$ are produced with the correct quantum
algebra SU$\left(  2,2\right)  =$SO$\left(  4,2\right)  $ in \textit{any
theory}, (for example bosonic spinors, or fermions $\psi^{M}$, or the list of
systems in section (\ref{s1}), or any other)\textit{ }then our formula
(\ref{solution}) gives the twistor transform for the corresponding degrees of
freedom of that theory. Those degrees of freedom appear as the building blocks
of $J^{MN}.$ So, the machinery proposed in this section contains some very
powerful tools.

\section{The Unifying SU$\left(  2,3\right)  $ Lie algebra}

The 2T-physics action (\ref{actionV}) offered the group SU$\left(  2,3\right)
$ as the most symmetric unifying property of the spinning particles for all
the systems listed in section (\ref{s1}), including twistors. Here we discuss
how this fundamental underlying structure governs and simplifies the quantum theory.

We examine the SU$\left(  2,3\right)  $ charges $\mathcal{J}_{A}^{~B},\hat
{J}_{0},j_{A},\bar{j}^{A}$ given in (\ref{conserved},\ref{J},\ref{J0}). Since
these are gauge invariant under all the gauge symmetries (\ref{locals}) they
are physical quantities that should have the properties of the Lie
algebra\footnote{Even when $j_{A}$ is not a conserved charge when the
U$\left(  1\right)  $ constraint is imposed, its commutation rules are still
the same in the covariant quantization approach, independently than the
constraint.} of SU$\left(  2,3\right)  $ in all the systems listed in section
(\ref{s1}). Using covariant quantization we construct the quantum version of
all these charges in terms of twistors. By using the general quantum twistor
transform of the previous section, these charges can also be written in terms
of the quantized spin and phase space degrees of freedom of any of the
relevant systems.

The twistor expressions for $\hat{J}_{0},\mathcal{J}_{A}^{~B}$ are already
given in Eqs.(\ref{hhat},\ref{JZZ})%
\begin{equation}
\hat{J}_{0}=\frac{1}{2}\left(  Z_{A}\bar{Z}^{A}+\bar{Z}^{A}Z_{A}\right)
,\;\;\mathcal{J}_{A}^{~B}=Z_{A}\bar{Z}^{B}-\frac{\hat{J}_{0}+2}{4}\delta
_{A}^{~B}. \label{definitions}%
\end{equation}
We have seen that at the classical level $\left(  j_{A}\right)  _{classical}%
=\sqrt{J_{0}}Z_{A}$ and now we must figure out the quantum version
$j_{A}=\sqrt{\hat{J}_{0}+\alpha}Z_{A}$ that gives the correct SU$\left(
2,3\right)  $ closure property%
\begin{equation}
\left[  j_{A},\bar{j}^{B}\right]  =\mathcal{J}_{A}^{~B}+\frac{5}{4}\hat{J}%
_{0}\delta_{A}^{~B}. \label{jj}%
\end{equation}
The coefficient $\frac{5}{4}$ is determined by consistency with the Jacobi
identity $\left[  \left[  j_{A},\bar{j}^{B}\right]  ,j_{C}\right]  +\left[
\left[  \bar{j}^{B},j_{C}\right]  ,j_{A}\right]  +\left[  \left[  j_{C}%
,j_{A},\right]  ,\bar{j}^{B}\right]  =0,$ and the requirement that the
commutators of $j_{A}$ with $\mathcal{J}_{A}^{~B},\hat{J}_{0}$ be just like
those of $Z_{A}$ given in Eqs.(\ref{commsJZ},\ref{commsJ1}), as part of the
SU$\left(  2,3\right)  $ Lie algebra. So we carry out the computation in
Eq.(\ref{jj}) as follows%
\begin{align}
\left[  j_{A},\bar{j}^{B}\right]   &  =\sqrt{\hat{J}_{0}+\alpha}Z_{A}\bar
{Z}^{B}\sqrt{\hat{J}_{0}+\alpha}-\bar{Z}^{B}\sqrt{\hat{J}_{0}+\alpha}%
\sqrt{\hat{J}_{0}+\alpha}Z_{A}\\
&  =\left(  \hat{J}_{0}+\alpha\right)  Z_{A}\bar{Z}^{B}-\left(  \hat{J}%
_{0}+\alpha-1\right)  \bar{Z}^{B}Z_{A}\label{commRule}\\
&  =\left(  \hat{J}_{0}+\alpha-1\right)  \left[  Z_{A},\bar{Z}^{B}\right]
+Z_{A}\bar{Z}^{B}\\
&  =\delta_{A}^{~B}\left(  \hat{J}_{0}+\alpha-1+\frac{\hat{J}_{0}+2}%
{4}\right)  +\mathcal{J}_{A}^{~B} \label{commRule2}%
\end{align}
To get (\ref{commRule}) we have used the properties $Z_{A}f\left(  \hat{J}%
_{0}\right)  =f\left(  \hat{J}_{0}+1\right)  Z_{A}$ and $\bar{Z}^{B}f\left(
\hat{J}_{0}\right)  =f\left(  \hat{J}_{0}-1\right)  \bar{Z}^{B}$ for any
function $f\left(  \hat{J}_{0}\right)  .$ These follow from the commutator
$\left[  \hat{J}_{0},Z_{A}\right]  =-Z_{A}$ written in the form $Z_{A}\hat
{J}_{0}=\left(  \hat{J}_{0}+1\right)  Z_{A}$ which is used repeatedly, and
similarly for $\bar{Z}^{B}.$ To get (\ref{commRule2}) we have used $\left[
Z_{A},\bar{Z}^{B}\right]  =\delta_{A}^{~B}$ and then used the definitions
(\ref{definitions}). By comparing (\ref{commRule2}) and (\ref{jj}) we fix
$\alpha=1/2.$ Hence the correct quantum version of $j_{A}$ is%
\begin{equation}
j_{A}=\sqrt{\hat{J}_{0}+\frac{1}{2}}Z_{A}=Z_{A}\sqrt{\hat{J}_{0}-\frac{1}{2}%
}~.
\end{equation}
The second form is obtained by using $Z_{A}f\left(  \hat{J}_{0}\right)
=f\left(  \hat{J}_{0}+1\right)  Z_{A}.$

Note the following properties of the $j_{A},\bar{j}^{A}$
\begin{align}
\bar{j}^{A}j_{A}  &  =\sqrt{\hat{J}_{0}-\frac{1}{2}}\bar{Z}Z\sqrt{\hat{J}%
_{0}-\frac{1}{2}}=\left(  \hat{J}_{0}-\frac{1}{2}\right)  \left(  \hat{J}%
_{0}-2\right) \label{jj1}\\
j_{A}\bar{j}^{B}  &  =\sqrt{\hat{J}_{0}+\frac{1}{2}}Z_{A}\bar{Z}^{B}\sqrt
{\hat{J}_{0}+\frac{1}{2}}=\left(  \hat{J}_{0}+\frac{1}{2}\right)  \left(
\mathcal{J+}\frac{1}{4}\left(  \hat{J}_{0}+2\right)  \right)  \label{jj2}%
\end{align}
which will be used below.

With the above arguments we have now constructed the quantum version of the
SU$\left(  2,3\right)  $ charges written as a $5\times5$ traceless matrix
\begin{align}
\hat{J}_{2,3}  &  =\left(  g^{-1}\left(
\genfrac{}{}{0pt}{}{\mathcal{L}}{0}%
\genfrac{}{}{0pt}{}{0}{0}%
\right)  g\right)  _{\text{quantum}}=\left(
\genfrac{}{}{0pt}{}{\mathcal{J+}\frac{1}{4}\hat{J}_{0}}{-\bar{j}}%
\genfrac{}{}{0pt}{}{j}{-\hat{J}_{0}}%
\right) \\
&  =\left(
\begin{array}
[c]{cc}%
Z_{A}\bar{Z}^{B}-\frac{1}{2}\delta_{A}^{~B} & \sqrt{\hat{J}_{0}+\frac{1}{2}%
}Z_{A}\\
-\bar{Z}^{B}\sqrt{\hat{J}_{0}+\frac{1}{2}} & -\hat{J}_{0}%
\end{array}
\right)  ,\;
\end{align}
with $\hat{J}_{0},\mathcal{J}$ given in Eq.(\ref{definitions}).

At the classical level, the square of the matrix $J_{2,3}$ vanishes since
$\mathcal{L}^{2}=0$ as follows
\begin{equation}
\left[  \left(  J_{2,3}\right)  ^{2}\right]  _{classical}=\left(
g^{-1}\left(
\genfrac{}{}{0pt}{}{\mathcal{L}}{0}%
\genfrac{}{}{0pt}{}{0}{0}%
\right)  g\right)  \left(  g^{-1}\left(
\genfrac{}{}{0pt}{}{\mathcal{L}}{0}%
\genfrac{}{}{0pt}{}{0}{0}%
\right)  g\right)  =g^{-1}\left(
\genfrac{}{}{0pt}{}{\mathcal{L}^{2}}{0}%
\genfrac{}{}{0pt}{}{0}{0}%
\right)  g=0.
\end{equation}
At the quantum level we find the following non-zero result which is SU$\left(
2,3\right)  $ covariant%
\begin{align}
\left(  \hat{J}_{2,3}\right)  ^{2}  &  =\left(
\begin{array}
[c]{cc}%
Z\bar{Z}-\frac{1}{2} & \sqrt{\hat{J}_{0}+\frac{1}{2}}Z\\
-\bar{Z}\sqrt{\hat{J}_{0}+\frac{1}{2}} & -\hat{J}_{0}%
\end{array}
\right)  ^{2}\label{su23}\\
&  =-\frac{5}{2}\left(  \hat{J}_{2,3}\right)  -1. \label{su233}%
\end{align}
By repeatedly using the same equation we can compute all powers $\left(
\hat{J}_{2,3}\right)  ^{n},$ and by taking traces we obtain the Casimir
eigenvalues of the SU$\left(  2,3\right)  $ representation. For example the
quadratic Casimir is%
\begin{equation}
Tr\left(  \left(  \hat{J}_{2,3}\right)  ^{2}\right)  =-5.
\end{equation}
Written out in terms of the charges, Eq.(\ref{su233}) becomes%
\begin{equation}
\left(
\genfrac{}{}{0pt}{}{\mathcal{J+}\frac{1}{4}\hat{J}_{0}}{-\bar{j}}%
\genfrac{}{}{0pt}{}{j}{-\hat{J}_{0}}%
\right)  ^{2}=-\frac{5}{2}\left(
\genfrac{}{}{0pt}{}{\mathcal{J+}\frac{1}{4}\hat{J}_{0}}{-\bar{j}}%
\genfrac{}{}{0pt}{}{j}{-\hat{J}_{0}}%
\right)  -1.
\end{equation}
Collecting terms in each block we obtain the following relations among the
gauge invariant charges $\mathcal{J},\hat{J}_{0},j,\bar{j}$%
\begin{align}
\left(  \mathcal{J+}\frac{1}{4}\hat{J}_{0}\right)  ^{2}-j\bar{j}+\frac{5}%
{2}\left(  \mathcal{J+}\frac{1}{4}\hat{J}_{0}\right)  +1  &  =0,\\
\left(  \mathcal{J+}\frac{1}{4}\hat{J}_{0}\right)  j-j\hat{J}_{0}+\frac{5}%
{2}j  &  =0,\\
-\bar{j}j+\left(  \hat{J}_{0}\right)  ^{2}-\frac{5}{2}\hat{J}_{0}+1  &  =0.
\end{align}
Combined with the information in Eq.(\ref{jj2}) the first equation is
equivalent to the master quantum equation (\ref{JJtoJ2}). After using
$j\hat{J}_{0}=\hat{J}_{0}j+j,$ the second equation is equivalent to the
eigenvalue equation (\ref{eigenvalue}) whose solution is the quantum twistor
transform (\ref{solution}). The third equation is identical to (\ref{jj1}).

Hence the SU$\left(  2,3\right)  $ quantum property $\left(  \hat{J}%
_{2,3}\right)  ^{2}=-\frac{5}{2}\left(  \hat{J}_{2,3}\right)  -1,$ or
equivalently $\left(  \hat{J}_{2,3}+2\right)  \left(  \hat{J}_{2,3}+\frac
{1}{2}\right)  =0,$ governs the quantum dynamics of all the sytems listed in
section (\ref{s1}) and captures all of the physical information, twistor
transform, and dualities as a property of a fixed SU$\left(  2,3\right)  $
representation whose generators satisfy the given constraint. This is a
remarkable simple unifying description of a diverse set of spinning systems,
that shows the existence of the sophisticated higher structure SU$\left(
2,3\right)  $ for which there was no clue whatsoever from the point of view of 1T-physics.

\section{Future Directions}

One can consider several paths that generalizes our discussion, including the following.

\begin{itemize}
\item It is straightforward to generalize our theory by replacing SU$\left(
2,3\right)  $ with the supergroup SU$\left(  2,\left(  2+n\right)  |N\right)
.$ This generalizes the spinor $V_{A}$ to $V_{A}^{a}$ where $a$ labels the
fundamental representation of the supergroup SU$\left(  n|N\right)  .$ The
case of $N=0$ and $n=1$ is what we discussed in this paper. The case of $n=0$
and any $N$ relates to the superparticle with $N$ supersymmetries (and all its
duals) discussed in \cite{super2t} and in \cite{twistorBP1}\cite{twistorLect}.
The massless particle gauge is investigated in \cite{fedo3}, but the other
cases listed in section (\ref{s1}) remain so far unexplored. The general model
has global symmetry SU$\left(  2,2\right)  \times$SU$\left(  n|N\right)
\times$U$\left(  1\right)  \subset\lbrack$SU$\left(  2,\left(  2+n\right)
|N\right)  ]_{R}$ if a U$\left(  1\right)  $ gauging is included, or the full
global symmetry [SU$\left(  2,\left(  2+n\right)  |N\right)  ]_{R}$ in its
high spin version. It also has local gauge symmetries that include bosonic \&
fermionic kappa symmetries embedded in [SU$\left(  2,\left(  2+n\right)
|N\right)  ]_{L}$ as well as the basic Sp$\left(  2,R\right)  $ gauge
symmetry. The gauge symmetries insure that the theory has no negative norm
states. In the massless particle gauge, this model corresponds to
supersymmetrizing spinning particles rather than supersymmetrizing the zero
spin particle. The usual R-symmetry group in SUSY is replaced here by
SU$\left(  n|N\right)  \times$U$\left(  1\right)  .$ For all these cases with
non-zero $n,N$, the 2T-physics and twistor formalisms unify a large class of
new 1T-physics systems and establishes dualities among them.

\item One can generalize our discussion in 4+2 dimensions, including the
previous paragraph, to higher dimensions. The starting point in 4+2 dimensions
was SU$\left(  2,2\right)  =$SO$\left(  4,2\right)  $ embedded in
$g=$SU$\left(  2,3\right)  .$ For higher dimensions we start from SO$\left(
d,2\right)  $ and seek a group or supergroup that contains SO$\left(
d,2\right)  $ in the spinor representation. For example for 6+2 dimensions,
the starting point is the 8$\times8$ spinor version of SO$\left(  8^{\ast
}\right)  =$SO$\left(  6,2\right)  $ embedded in $g=$SO$\left(  9^{\ast
}\right)  =$SO$\left(  6,3\right)  $ or $g=$SO$\left(  10^{\ast}\right)
=$SO$\left(  6,4\right)  .$ The spinor variables in 6+2 dimensions $V_{A}$
will then be the spinor of SO$\left(  8^{\ast}\right)  =$SO$\left(
6,2\right)  $ parametrizing the coset SO$\left(  9^{\ast}\right)  /$SO$\left(
8^{\ast}\right)  $ (real spinor) or SO$\left(  10^{\ast}\right)  /$SO$\left(
8^{\ast}\right)  \times$SO$\left(  2\right)  $ (complex spinor). This can be
supersymmetrized. The pure superparticle version of this program for various
dimensions is discussed in \cite{twistorBP1}\cite{twistorLect}, where all the
relevant supergroups are classified. That discussion can now be taken further
by including bosonic variables embedded in a supergroup as just outlined in
the previous item. As explained before \cite{twistorBP1}\cite{twistorLect}, it
must be mentioned that when $d+2$ exceeds $6+2$ it seems that we need to
include also brane degrees of freedom in addition to particle degrees of
freedom. Also, even in lower dimensions, if the group element $g$ belongs to a
group larger than the minimal one \cite{twistorBP1}\cite{twistorLect}, extra
degrees of freedom will appear.

\item The methods in this paper overlap with those in \cite{2tAdSs} where a
similar master quantum equation technique for the supergroup SU$\left(
2,2|4\right)  $ was used to describe the spectrum of type-IIB supergravity
compactified on AdS$_{5}\times$S$^{5}.$ So our methods have a direct bearing
on $M$ theory. In the case of \cite{2tAdSs} the matrix insertion $\left(
\genfrac{}{}{0pt}{}{\mathcal{L}}{0}%
\genfrac{}{}{0pt}{}{0}{0}%
\right)  $ in the 2T-physics action was generalized to $\left(
\genfrac{}{}{0pt}{}{\mathcal{L}_{\left(  4,2\right)  }}{0}%
\genfrac{}{}{0pt}{}{0}{\mathcal{L}_{\left(  6,0\right)  }}%
\right)  $ to describe a theory in 10+2 dimensions. This approach to higher
dimensions can avoid the brane degrees of freedom and concentrate only on the
particle limit. Similar generalizations can be used with our present better
develped methods and richer set of groups mentioned above to explore various
corners of $M$ theory.

\item One of the projects in 2T-physics is to take advantage of its flexible
gauge fixing mechanisms in the context of 2T-physics field theory. Applying
this concept to the 2T-physics version of the Standard Model
\cite{2tstandardM} will generate duals to the Standard Model in 3+1
dimensions. The study of the duals could provide some non-perturbative or
other physical information on the usual Standard Model. This program is about
to be launched in the near future \cite{ibquelin}. Applying the twistor
techniques developed here to 2T-physics field theory should shed light on how
to connect the Standard Model with a twistor version. This could lead to
further insight and to new computational techniques for the types of twistor
computations that proved to be useful in QCD \cite{witten2}\cite{cachazo}.

\item Our new models and methods can also be applied to the study of high spin
theories by generalizing the techniques in \cite{vasiliev} which are closely
related to 2T-physics. The high spin version of our model has been discussed
in many of the footnotes, and can be supersymmetrized and written in higher
dimensions as outlined above in this section. The new ingredient from the 2T
point of view is the bosonic spinor $V_{A}$ and the higher symmetry, such as
SU$\left(  2,3\right)  $ and its generalizations in higher dimensions or with
supersymmetry. The massless particle gauge of our theory in 3+1 dimensions
coincides with the high spin studies in \cite{fedo1}-\cite{fedo4}. Our theory
of course applies broadly to all the spinning systems that emerge in the other
gauges, not only to massless particles. The last three sections on the quantum
theory discussed in this paper would apply also in the high spin version of
our theory. The more direct 4+2 higher dimensional quantization of high spin
theories including the spinor $V_{A}$ (or its generalizations $V_{A}^{a}$) is
obtained from our SU$\left(  2,3\right)  $ quantum formalism in the last section.

\item One can consider applying the bosonic spinor that worked well in the
particle case to strings and branes. This may provide new string backgrounds
with spin degrees of freedom other than the familiar Neveu-Schwarz or
Green-Schwarz formulations that involve fermions.
\end{itemize}

More details and applications of our theory will be presented in a companion
paper \cite{twistorBO}.\bigskip

We gratefully acknowledge discussions with S-H. Chen, Y-C. Kuo, and G. Quelin.

\end{document}